\documentclass[11pt,a4paper]{article}
\title{ Quantum entanglement in coupled lossy waveguides using \\ $SU(2)$ and $SU(1,1)$ Thermo-algebras. }
\def\be{\begin{equation}}
\def\ee{\end{equation}}
\def\bea{\begin{eqnarray}}
\def\eea{\end{eqnarray}}

\def\dag{\dagger}

\usepackage{amsmath,amsfonts,amssymb}
\usepackage{graphicx}
\usepackage{epstopdf}
\author{ Naveen Kumar M.$^a$ \footnote{Email: naveenhere928@gmail.com}
K. V. S. Shiv Chaitanya$^b$ \footnote{ Email: chaitanya@hyderabad.bits-pilani.ac.in}
Bindu A. Bambah$^a$  \footnote{\small{Email:  bbsp@uohyd.ernet.in}}.
\\
 $^a$ \small School of Physics,
 University of Hyderabad,  Hyderabad, Telangana-500046, India. \\
 $^b$ \small Department of Physics, BITS-Pilani, Hyderabad Campus, Hyderabad, Telangana-500072, India.
 }
\begin{document}
\maketitle
\begin{abstract}
\begin{center}
{\small }
\end{center}
In this paper, the master equation for the coupled lossy waveguides  is solved using  the thermofield dynamics(TFD) formalism.
This  formalism allows  the use of the  underlying symmetry algebras $SU(2)$ and $SU(1,1)$,  associated with the Hamiltonian  of  the coupled lossy waveguides,
to compute entanglement and decoherence as a function of time for various input  states such as NOON states and thermal states.

\end{abstract}

\section{Introduction}
 Recently, there has been a lot of interest in studying entanglement using coupled waveguides\cite{Christ, Longhi}. 
Specially designed  photonic waveguides have provided a laboratory tool for analyzing coherent quantum phenomena and
have a possible application in quantum computation\cite{Longhi1}.
The entanglement between waveguide
modes is at the heart of many of these experiments and has been widely studied\cite{Agarwal}. Using coupled silica waveguides Politi et.al.\cite{Politi} 
have reported control of multiphoton entanglement directly on-chip  and demonstrated integrated quantum metrology opening the way for new  quantum technologies.
They have been able to generated two and four photon NOON states on the chip and observed quantum interference,
which further enhances the capabilities for quantum interference and quantum computing.
Among various types of entangled states, NOON states are special with two orthogonal states in maximal super position
thus enhancing their use in quantum information processing\cite{Rozema}.

For the  efficient use of these waveguides in the field of quantum information, the generated entanglement should not decohere with time\cite{Zurek}. 
It is well known that  losses have a substantial effect on the wave guides. Therefore, the time evolution of  the entanglement  is of interest in the context of quantum information processing using lossy waveguides.
The entanglement between waveguide modes and how loss affects this entanglement has recently been studied by Rai et.al.\cite{Agarwal},
by using a quantum Liouville equation. In this paper, we approach this problem from the viewpoint of thermofield dynamics\cite{Umezawa}.
This formalism  has the advantage of 
 solving the master equation exactly  for  both pure and mixed states, converting thermal averages into quantum mechanical expectation values, by doubling the Hilbert space.
The formalism thus makes it simpler to calculate the effects of noise and decoherence in the coupled two mode waveguide system.
We look at the effect of different type of input states 
and show the efficiency of the states for quantum information theory.

\subsection{Thermofield dynamics}
Thermofield dynamics(TFD)\cite{Umezawa, Laplae, Takahashi, Ojima,  Umezawa2} is a finite temperature field theory. It has been applied to many branches of high
energy physics and many-body systems. TFD has been used to solve the master equation, which helps in studying the temporal evolution of 
entanglement. In particular, it has been used to study entanglement in the presence of a Kerr medium using an SU(1,1) disentanglement theorem
for  arbitrary initial conditions by reducing it to a Schrodinger-like equation\cite{Chaturvedi, Chaturvedi2}.
Thus, all the techniques available to solve the Schrodinger equation can be used to solve the master equation. We  derive the effect of losses giving rise
to decoherence by solving  the master equation exactly, using TFD, and then compute  the decoherence and entanglement properties of some two mode waveguide systems.
\\
The basic formalism of TFD is as follows:
corresponding to the  the creation and annihilation operators
$a^\dag$ and $a$, which act on the physical space $\mathcal{H}$, we introduce  "thermal or tildian"  operators  $\tilde{a}^\dag$ and $\tilde{a}$,
which  act on an augmented(fictitious) space $ \mathcal{\tilde{H}}$\cite{Shanta, Shanta2, Chaturvedi3}.  
The operators $a$ and $a^\dag$ commute with $\tilde{a}$ and $\tilde{a}^\dag$ and
the sets of basis vectors  $\{|n>\}$ and $\{ |\tilde{n}>\}$  span the Hilbert spaces $\mathcal{H}$ and  $\mathcal{\tilde{H}}$, respectively,
and  $\{ |n> \times |\tilde{m}>\}=|n,\tilde{m​}>$ are the basis operators of the doubled Hilbert space.
The Identity operator is $|I>= \sum |n>\otimes |\tilde{n}>=\sum |n,\tilde{n}>$.
We define a temperature dependent `` thermal vacuum state'', 
\be |0(\beta)> = e^{-iG(\theta)}|0,\tilde{0}>; \hspace{0.7cm} with, \hspace{0.3cm}
G(\theta(\beta))=-i \theta(\beta)(\tilde{a}a - \tilde{a}^\dag a^\dag), \label{thermalstate}\ee
where,  $\tanh\theta(\beta) = e^{-\beta \omega/2} $ and the corresponding thermal  number distribution  is $\bar{n}(\beta)=1/(e^{\beta\omega}-1) $. The
statistical average of any operator  is equal to the vacuum expectation value of the operator with respect to the thermal vacuum.
In particular,  the density operator $\rho$ acting on a Hilbert space $\mathcal{H}$ is a state vector $|\rho^\alpha >$, $\frac{1}{2} \le \alpha \le 1 $ in 
the extended Hilbert space,   so that averages of operators with respect to $\rho$ are expressed as  scalar products, \vspace{-0.5cm}

\be <A> = Tr(A \rho ) = < \rho^{1-\alpha} | A |\rho^\alpha >,\ee

where ,the state $|\rho^\alpha > = \hat{\rho}^\alpha|I> $ and $\hat{\rho}^\alpha =\rho^\alpha \otimes I $. Although, in some applications of TFD the $\alpha=\frac{1}{2}$ formalism is used, in this paper
we work with the $\alpha=1$ formalism such that $<A><I|A|\rho>=<A|\rho>$

In TFD,  the master equation is  

\be \frac{\partial}{\partial t} |\rho(t)> = -i \hat{H} |\rho> \label{mste} ,\ee

where, $|\rho >$ is a vector in the extended Hilbert space $\mathcal{H} \otimes \mathcal{H}^* $ and $ -i\hat{H} = i(H - \tilde{H})+L$, L is the Liouville term. 
Thus, the problem of solving master equation is reduced to
solving a Schrodinger like equation. Symmetries  associated with the Hamiltonian(such as $SU(2)$ and $SU(1,1)$
symmetries) can be  exploited to  this equation. This makes the study of entanglement in lossy systems very tractable.
\section{Coupled waveguide system}
In optical communications, coupled waveguides are used as a  transmission medium.
The  linear coupling between two wave guides is used to transfer the power one wave guide to another.
To study the decoherence and entanglement properties of the two coupled waveguides, the model of  Rai and Agarwal\cite{Agarwal} is used.

The Hamiltonian described by,  \vspace{-0.3cm}

\be H =\hbar \omega (a^{\dag} a+ b^{\dag} b) + \hbar J(a^{\dag} b + b^{\dag} a) \label{hmltn}\ee
Where mode ‘a’ corresponds to first wave guide and mode ‘b’ corresponds to the second waveguide.
The mode ‘a’ and ‘b’ obey bosonic commutation relations.
The evanescent coupling in terms of distance between the two wave guide is given by ‘J’.
The density operator has a time evolution given by 
\be \frac{\partial}{\partial t} \rho = -\frac{i}{\hbar}[H,\rho] \label{qle0}\ee
In the presence of a damping term(system reservoir interaction), the evolution equations are governed by the Liouville equation
\be  \frac{\partial}{\partial t} \rho  = -\frac{i}{\hbar}[H,\rho]+ \mathcal{L} \rho, \label{qle}\ee
where,
 \be \mathcal{L} \rho = - \gamma (a^{\dag} a \rho - a \rho a^{\dag} + a^{\dag} \rho a + \rho a^{\dag} a 
 + b^{\dag} b \rho - b \rho b^{\dag} + b^{\dag} \rho b + \rho b^{\dag} b)\label{lrho}\ee
where ’$\gamma$’ is  dissipation in the material of the waveguide.

To study the entanglement and decoherence properties as function of time for the coupled wave guide system,
we solve the exactly master equation(\ref{lrho}) using TFD. This allows us to study the response to the coupling of different input states, such as,  number,
NOON and thermal states, in the coupled waveguide system.
In particular,  in the absence of damping, an input vacuum state evolves into a two mode SU(2) coherent state.
In the presence of damping,  the vacuum state evolves into a two mode squeezed state and a thermal state into a thermal squeezed state.

 \subsection{Two Coupled waveguides  without damping($\gamma = 0$)}

 The time evolution of the density operator corresponding to the two coupled waveguides  without damping  consisting of fields in the modes
$a$ and b from eqs.(\ref{hmltn}, \ref{qle0}) is determined  by,  \vspace{-0.4cm}
\be \dot{\rho} =  -i\omega (a^\dag a \rho  + b^\dag b \rho - \rho a^\dag a -\rho  b^\dag b) -i J (a^\dag b\rho  + b^\dag a\rho -\rho a^\dag b -\rho b^\dag a) \label{qle0expand1} \ee
  \vspace{-0.1cm}
Now we apply the TFD-formalism(eq.(\ref{mste})),  by doubling the Hilbert space and get the Schrodinger type wave equation   \vspace{-0.5cm}

  \be |\dot{\rho}> = -i \hat{H}|\rho> \label{eqnofmtn0}\ee
  here, the Hamiltonian $\hat{H}$ is given by \vspace{-0.5cm}  \be \hat{H}  =  \omega (a^\dag a +b^\dag b - \tilde{a}^\dag \tilde{a}  - \tilde{b}^\dag \tilde{b})
  +  J(a^\dag b + b^\dag a - \tilde{a}^\dag \tilde{b} -  \tilde{b}^\dag \tilde{a}) \ee
  
  and can be  decoupled into non-tildian and tildian parts: $  \hat{H} = H-\tilde{H}, $
  
  where,   \vspace{-0.4cm}
  \be H =  \omega (a^\dag a +b^\dag b) +  J(a^\dag b + b^\dag a) \label{Hamiltonian1} \ee \vspace{-0.7cm} \be \hspace{0.5cm} \tilde{H} =  \omega (\tilde{a}^\dag \tilde{a}  + \tilde{b}^\dag \tilde{b})
  + J( \tilde{a}^\dag \tilde{b} + \tilde{b}^\dag \tilde{a}) \label{Hamiltonian2}\ee 
    \vspace{-0.3cm}
    
    Then the solution of eq.(\ref{eqnofmtn0}) is given by   \vspace{-0.2cm} \be |\rho(t)> = exp[-iHt] \otimes exp[-i\tilde{H}t])|\rho(0)> \label{densitystate} \ee
    where, $|\rho(0) >$ is an intial state in $\mathcal{H} \otimes \mathcal{\tilde{H}}. $
    
It is clear from the above that, the two Hamiltonians H and $\tilde{H}$ are independent. 
Hence, we can work with one of the Hamiltonians but since we are only interested in the physical states, we trace over the tilde states.

To calculate the decoherence and entanglement properties  the  master equation(\ref{qle0expand1}),is solved using the  underlying symmetries associated with the Hamiltonians(eqns(\ref{Hamiltonian1}, \ref{Hamiltonian2})) .
To see these  symmetries explicitly we define the following operators,
\vspace{-0.5cm}

\be  \hspace{0.3cm} L_+ =a^\dag b, \hspace{0.5cm} L_- = b^\dag a \hspace{0.5cm} \& \hspace{0.2cm}  L_3 = \frac{1}{2} (a^\dag a - b^\dag b) \label{SU(2)generators} , \ee
 \vspace{-0.2cm}
which satisfy the $SU(2)$ algebra,  \vspace{-0.5cm}

\be [L_3, L_\pm ] = \pm L_\pm \hspace{0.7cm} \& \hspace{0.7cm} [L_+, L_-] = 2 L_3 \label{SU(2)C-relations} \ee  with number operator, $\mathcal{N} = a^\dag a +b^\dag b $.

The Hamiltonian ,  in terms of the SU(2) generators,  is \vspace{-0.5cm}

 \be H = \omega \mathcal{N} + J(L_+ +L_-) . \ee
 
 The underlying symmetry of the Schrodinger like eq(\ref{eqnofmtn0}) is $SU(2) \otimes SU(2)$ and $|\rho(t)>$ 
is given by 
\vspace{-0.3cm}
\be |\rho(t)>  = e^{\alpha (t) L_+ - \alpha^*(t) L_-} |\rho(0)>; \hspace{0.7cm} here, \hspace{0.2cm}  \alpha (t) = iJt, \label{psi}. \ee 

Using the  disentanglement formula\cite{Perelomov, KazuyukiFujii}, and taking the initial state $|\rho(0)>$ as the vacuum state, the solution(\ref{psi}) is,
\vspace{-0.2cm}

\be |\rho(t)> = e^{\xi L_+} e^{\log(1+|\xi|^2) L_3} e^{- \xi^* L_-}|\rho(0)> \ee

where, $ \xi = \xi(\alpha(t)) = \frac{ \alpha(t) \tan(|\alpha(t)|)}{|\alpha(t)|}  \label{xi} $

The density matrix in the number state basis is 
\vspace{-0.4cm}

\begin{align} \nonumber |\rho(t)> &  =  \sum^{N}_{n_a, m_a = 0} C_{n_a,n_a} C^*_{m_a,m_a} |n_a,N-n_a><m_a,N-m_a|  \label{staterho}
\end{align}

where, \vspace{-0.5cm} \be C_{n_a,n_a} = \frac{\xi^{n_a}}{(1+|\xi|^2)^{N/2}}. \left(
\begin{array}{rcl}
N \\
 n_a \\
\end{array}
\right)^{\frac{1}{2}} ; \hspace{0.7cm} and \hspace{0.4cm}  N=n_a+n_b, \hspace{0.5cm} N=m_a+m_b. \ee

The entanglement properties are calculated by taking the partial transpose of $\rho$ and computing  the eigenvalues of the resulting  matrix\cite{Agarwal2} 
:

\be |\rho(t)_{PT}> = \sum^{N}_{n_a, m_a = 0} C_{n_a,n_a} C^*_{m_a, m_a} |n_a,N-m_a><m_a, N-n_a| \label{rhopt} \ee 
\be C_{n_a,n_a} C^*_{m_a,m_a} |n_a,N-m_a><m_a,N-n_a| +  C_{m_a,m_a} C^*_{n_a,n_a} |m_a,N-n_a><n_a,N-m_a|. \ee
The eigenvalues are 
\be \lambda_{n_an_a} =  \frac{N!}{(N-n_a)! n_a!} \sin^{2n_a}(Jt) \cos^{2N-2n_a}(Jt);  (for,  n_a=m_a) \label{lambda1}\ee \be
 \lambda_{n_am_a} = \pm \frac{N!}{\sqrt{(N-n_a)! (N-m_a)! n_a! m_a! }} \sin^{m_a+n_a}(Jt) \cos^{2N-m_a-n_a}(Jt); (for,  n_a\ne m_a).\label{lambda2} \ee 
 
 \subsubsection{Entanglement properties of the system}
 
 For a bipartite system, the entropy is defined as the von-Neumann entropy of the reduced density matrix traced with respect to one of the systems as
\vspace{-0.3cm}
\be S    = -Tr(\rho_a \log \rho_a)   = -\sum^N_{i=0} \lambda_i \log \lambda_i, \label{partialtrace} \ee such that $\rho_a = Tr_b(\rho)$ and
$ \{\lambda_i\} $ correspond to the set of eigenvalues of the reduced density operator.  For the general number state,
with the eigenvalues given by eqns(\ref{lambda1}, \ref{lambda2}) the entropy is  \vspace{-0.3cm}
 \begin{align} \nonumber S & =  -\sum^N_{n = 0} [ \frac{N!}{(N-n)! n!} \sin^{2n}(Jt) \cos^{2N-2n}(Jt)] 
 \\&  \hspace{1.2cm} \times \{ \log[\frac{N!}{(N-n)!  n!}] + (2n) \log[\sin(Jt)] + (2N-2n) \log[\cos(Jt)] \} \label{entropy}
\end{align}
 
 We can also  quantify the entanglement of the system by studying the time evolution for the logarithmic negativity\cite{Vidal}, 
 which is an easily computable measure of distillable entanglement.
 For a bipartite system described by the density matrix, $\rho$ the log negativity is, 
\be E_N(T) = \log_2\Vert\rho^T\Vert, \hspace{1cm} here, \hspace{0.3cm} \Vert\rho^T\Vert = (2N(\rho) + 1) \label{lognegativity} \ee 
where $\rho^T$ is the partial transpose of $\rho $ and the symbol $\Vert  \Vert$
denotes the trace norm. Also  $N(\rho)$ is the absolute value of the sum of all the negative eigenvalues of the partial transpose of $\rho$. The
log negativity is a non-negative quantity and a non-zero value of $E_N$ would mean that the state
is entangled.

Now,  we consider various cases of optical input states:
 
 {\bf \underline{Case-1}:} {\bf For two photon system as an input(i.e., $N=2$):}
 
 The entropy of entanglement of the two photon input state (both $|1,1>$ and $|2,0>$ ) is  \vspace{-0.3cm}
\be S  =- 4 \cos^4(Jt) \log[\cos(Jt)] - 2 \sin^2(Jt) \cos^2(Jt)  \log[\cos(Jt) \sin(Jt)] - 4 \sin^4(Jt)  \log[\sin(Jt)] \ee
and  is shown in dotdashed curve of fig.1(d). In this case, at time $t = 0$, we begin with a separable input
state and thus the value of Entropy of entanglement(S) is zero. Then the value of 'S' increases and attains a maximum value of 1.5 at $Jt=0.785212$.
Then decreases and eventually becomes equal to zero at $Jt=1.57061$. Thus the
state becomes disentangled at this point of time. At later times we see a periodic behavior and
the system gets entangled and disentangled periodically. We
do not see any interference effects.
 
The logarithmic negativity  of the various two photon states  is different for various two photon states, unlike the entropy.\\
 {\bf Case-1(a.)}  If we take the input state as $|\psi_{in}> = |1,1>$, the output state is
 
 $|\psi_{out}> = \alpha_1 |0,2> +\alpha_2|1,1> + \alpha_3 |2,0>.$
where, $ \alpha_1 = -i \sin(2Jt)/ \sqrt{2}, \alpha_2 = \cos(2Jt), \\
\alpha_3 = -i \sin(2Jt)/ \sqrt{2}.$ Then we can write the log negativity entanglement of this system(from eq.(\ref{lognegativity})) as, \vspace{-0.4cm}

\be E_N  = \log_2[1+2(-i \sqrt{2} \sin(2Jt) \cos(2Jt)-\sin^2(2Jt)/2)] \ee
which  is shown in thick curve of Fig.1(a).  At time $t = 0$, we begin with a separable input
state and thus the value of log negativity is zero. 
$E_N$ increases with time and attains a maximum value of 1.32875 for $Jt = 0.42879$, this is the maximally entangled state.
Further, for $Jt =0.785212$ we get the dips at $E_N = 1$(the coincidence rate of the output modes of the beam splitter will drop to zero,
when the identical input photons overlap perfectly in time), due to  Hong-Ou-Mandel
interference\cite{Hong} and for $Jt = 1.57061$, $E_N$ vanishes. At later times,
we see a periodic behavior, attributed to the inter-waveguide coupling(J).

\vspace{0.2cm}

{\bf Case-1(b.)} Now we take the input state as $|\psi_{in}> = |2,0>$, the output state is 

$|\psi_{out}> = \beta_1 |0,2> +\beta_2|1,1> + \beta_3 |2,0>$ 
here, $ \beta_1 = \cos^2(Jt), \beta_2 = - i \sin(2Jt)/\sqrt{2}, \\
\beta_3 = -\sin^2(Jt).$
The log negativity  for this system is , 
\be E_N = \log_2[1+2(-i \sqrt{2} \sin(Jt) \cos^3(Jt) + i \sqrt{2} \sin^3(Jt) \cos(Jt) - \sin^2(Jt) \cos^2(Jt))] \ee
and is shown in dotted curve of Fig.1(a).  In this case, $E_N$ increases and attains a maximum value of 1.32193 at $Jt=0.785212$,
then decreases and eventually becomes equal to zero at $Jt=1.57061$. Thus the
state becomes disentangled at this point of time. At later times we see a periodic behavior and
the system gets entangled and disentangled periodically. Unlike the earlier case for the $|1,1>$ input state, we
do not see any interference effects. Clearly the entanglement dynamics of
the states $|1,1>$ and $|2,0>$ are different.\\

{\bf Case-1(c.)} For two photon input NOON state:

$|\psi_{in}> = (|2,0> + |0,2>)/\sqrt{2},$  the output state will be, $|\psi_{out}> = a_1 |0,2> +a_2 |1,1> + a_3 |2,0>.$ 
 
 where, $a_1 = \cos(2Jt)/\sqrt{2} , a_2 = -i \sin(2Jt) , a_3 = \cos(2Jt)/\sqrt{2}. $
The logarithmic negativity  of  this state \vspace{-0.4cm}

\be E_N  =\log_2[1+ i \sqrt{2}  \sin(4Jt) + \cos^2(2Jt)]
\ee
is shown in thick curve of Fig.1(c). At time $t = 0$, we begin with an entangled $|2002>$ state as input and thus the value of log negativity is one. 
We see periodic dips due to the Hong-Ou-Mandel
interference, which is characteristic of NOON states. 
 $E_N$ increases with time and attains a maximum value of 1.32875 for $Jt = 0.356988$, which is the point of maximal entanglement.
Further, for $Jt = 0.785212$, $E_N$ vanishes(we will see later that as we increase the number of photons in the NOON state,
the entanglement does not vanish). At later times,
we see a periodic behavior, attributed to the inter-waveguide coupling(J).

\begin{figure}
\begin{center}
\includegraphics[width=3in,height=2.4in]{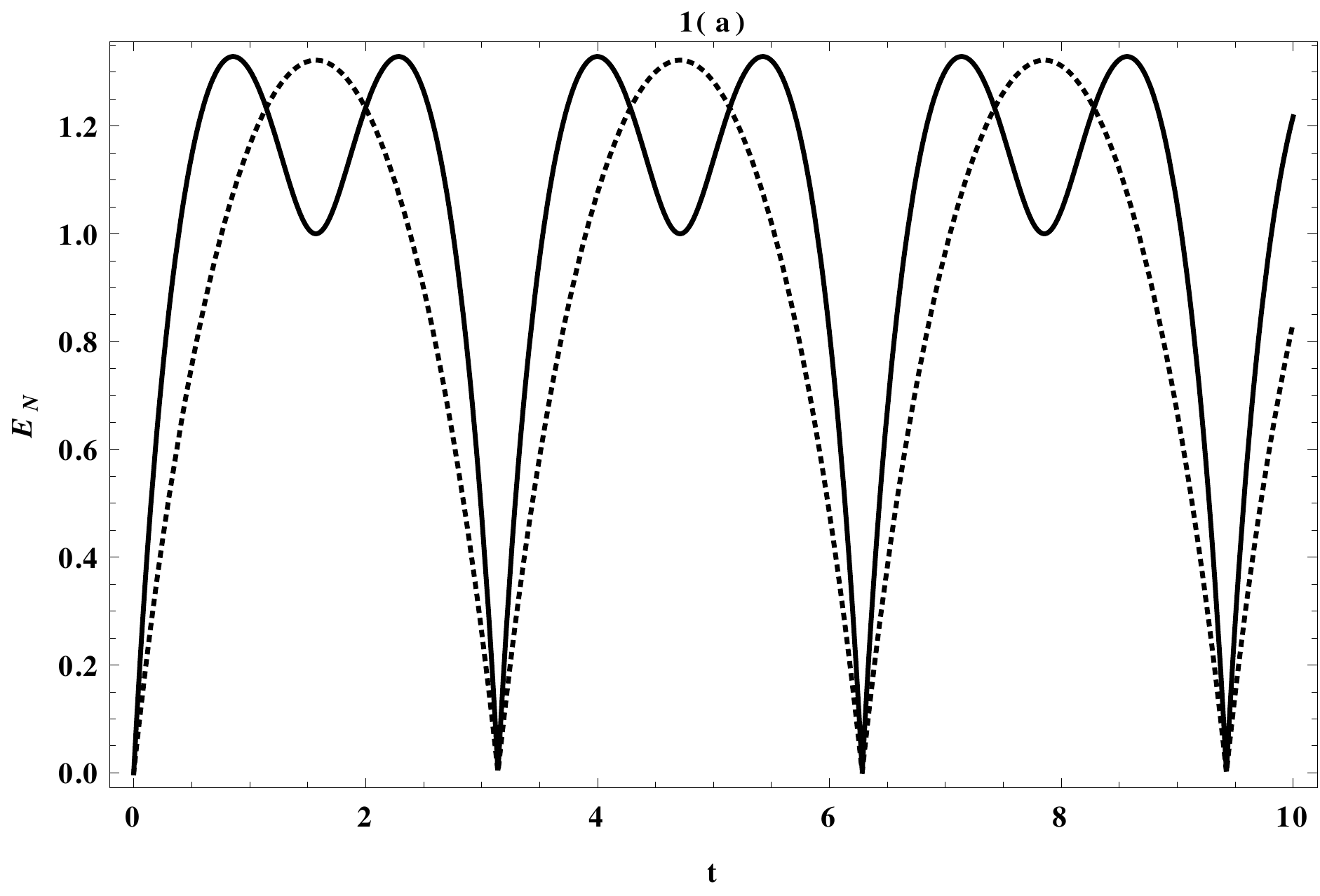} \hspace{0.2cm} \includegraphics[width=3in,height=2.4in]{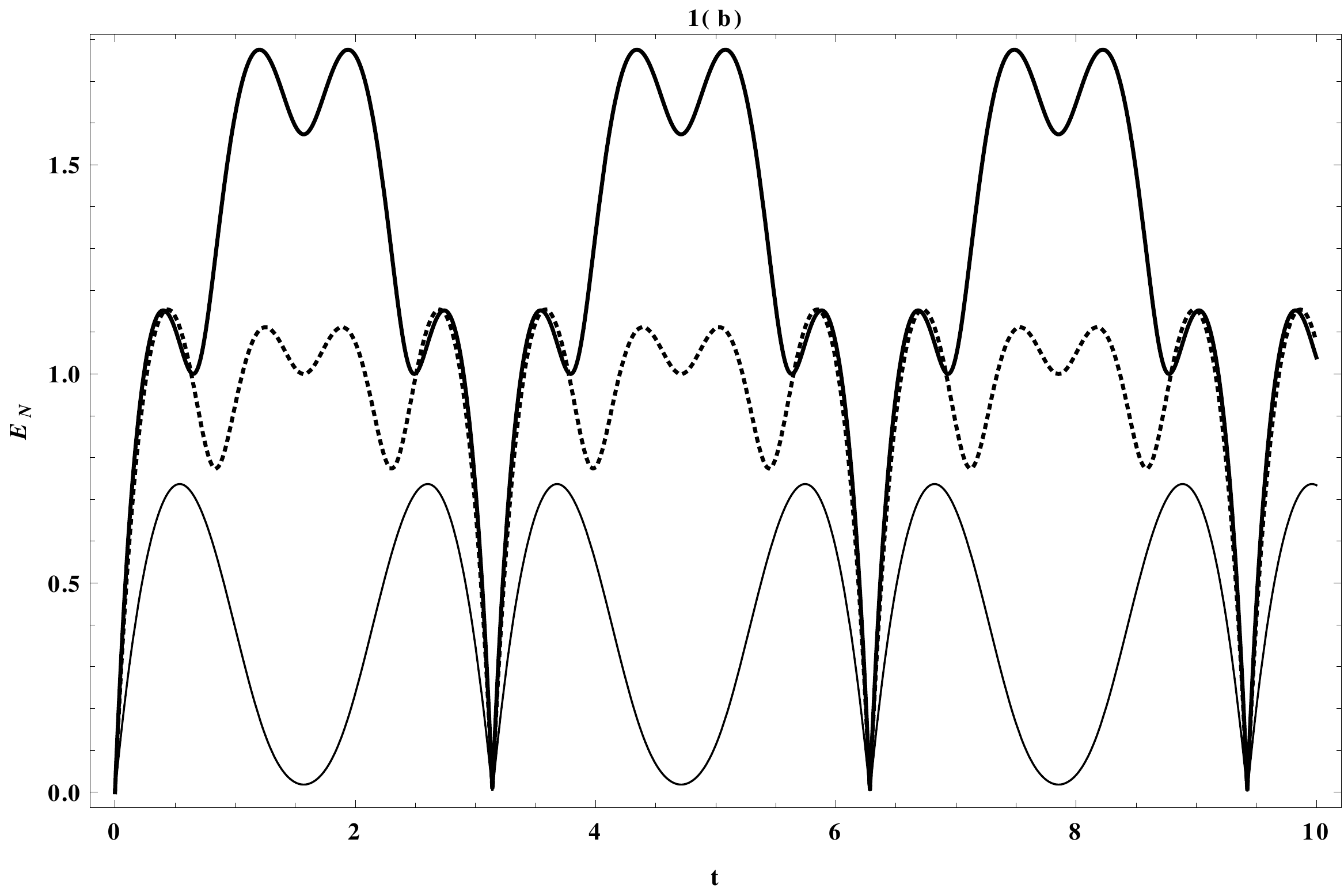} \\
\vspace{0.7cm}
\includegraphics[width=3.15in,height=2.0in]{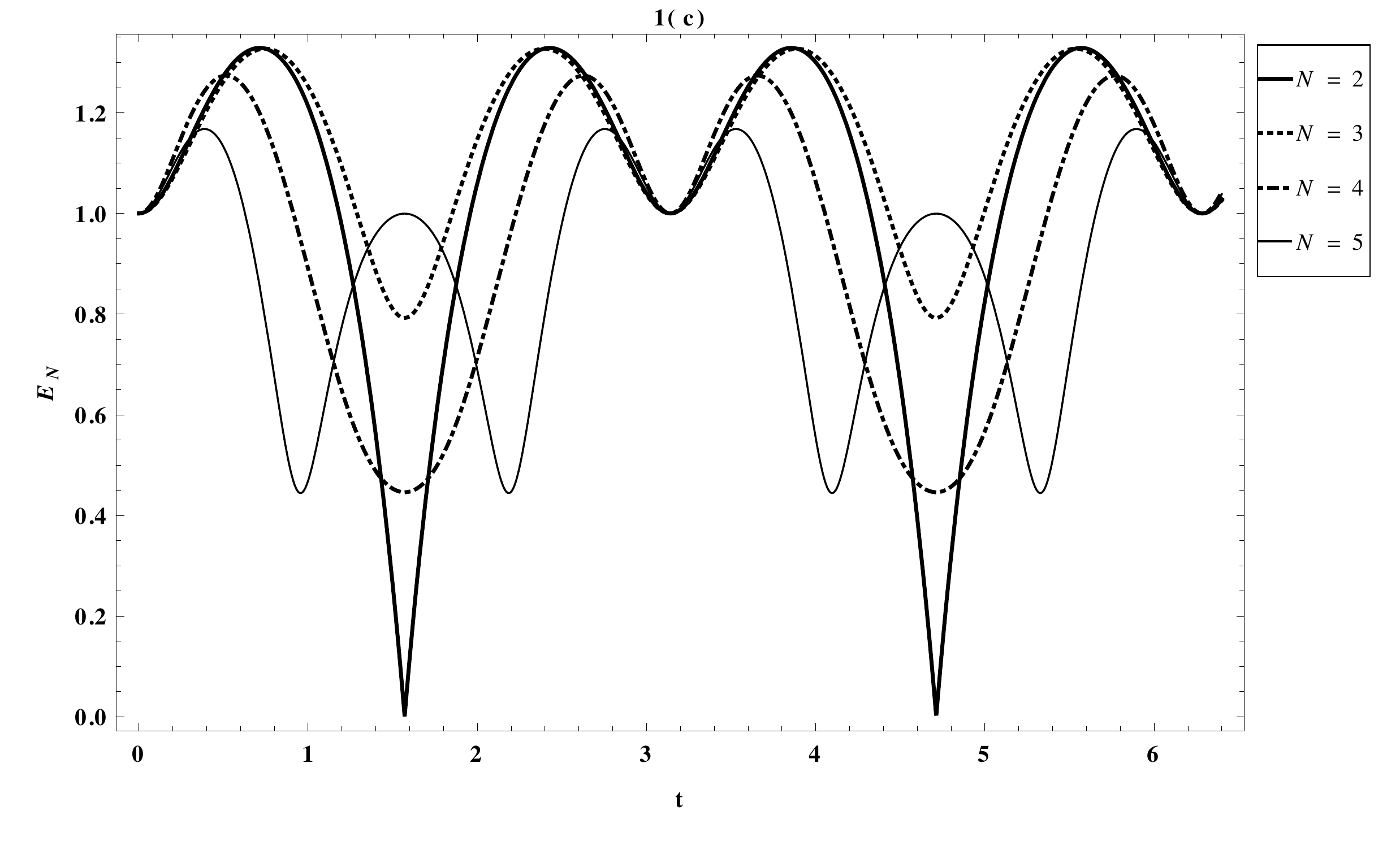} \hspace{0.1cm}  \includegraphics[width=3.1in,height=2.0in]{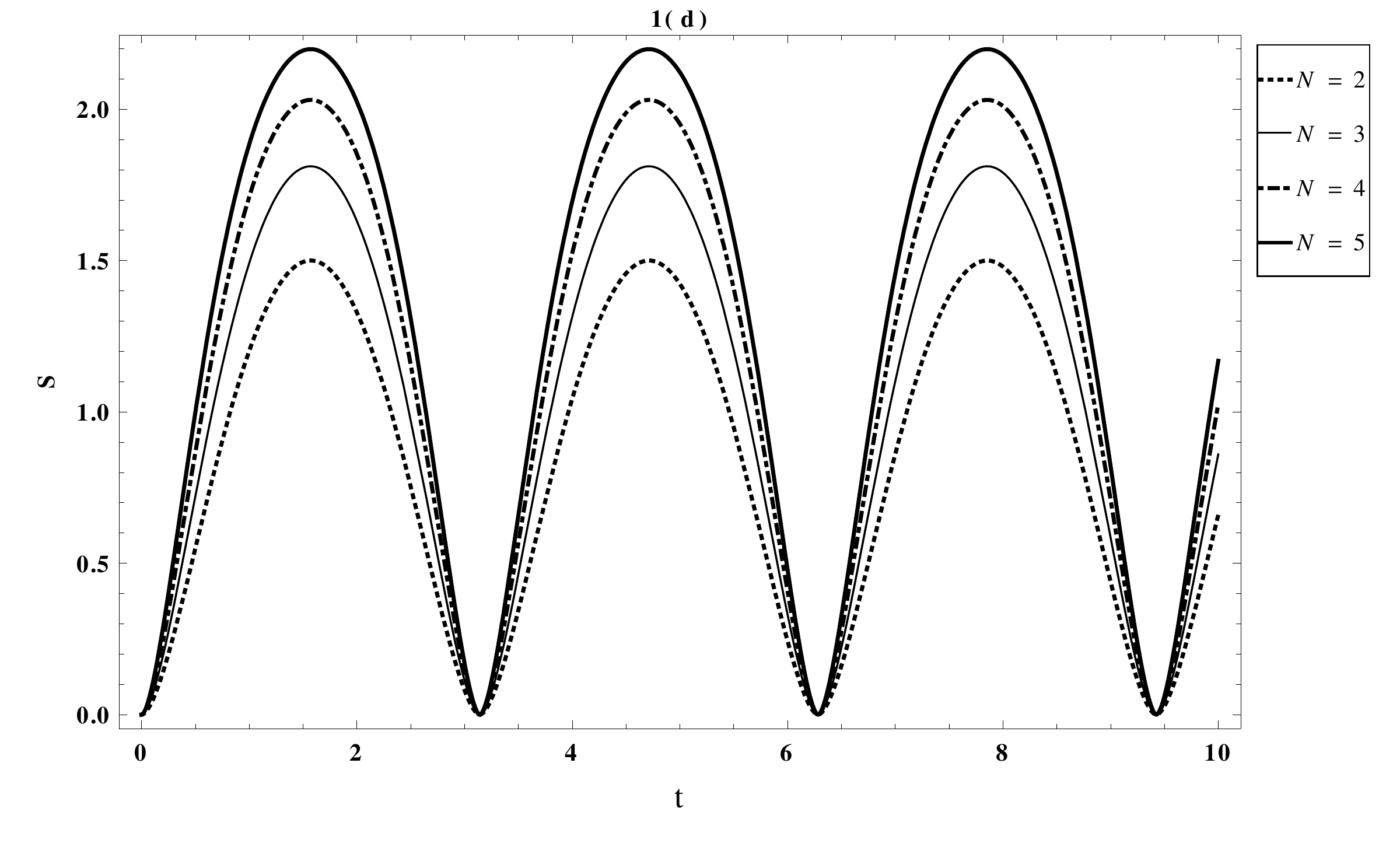} \\
\end{center}
\caption{(1a) Shows the time evolution of log negativity for a  two photon system: 
the thick curve shows the result for $|1,1>$ state and the dotted curve shows the result for the $|2,0>$ state.
(1b) shows the time evolution of log negativity for a four photon system: the thick curve shows the result for $|2,2>$ state, 
the dotted curve shows the result for $|3,1>$ state, and the thin curve shows the result for $|4,0>$ state.
(1c) shows the time evolution of log negativity for N photon NOON states (N=2,3,4,5).
(1d) shows the time evolution of entropy of entanglement(S) for N photon system(N=2,3,4,5).} \label{1}
\end{figure}

{\bf\underline{Case-2}:} {\bf For four photon system as an input(i.e.,$N=4$):}

The entropy of entanglement for four photon system is\\  \vspace{-0.5cm}
\begin{align} \nonumber  S  = &- 8 \cos^8(Jt)\log[\cos(Jt)]  - 8 \sin^8(Jt)  \log[\sin(Jt)] - 4 \sin^2(Jt) \cos^6(Jt)  \log[4 \sin^2(Jt)  \cos^6(Jt)] 
\\& - 6 \sin^4(Jt) \cos^4(Jt)  \log[6 \cos^4(Jt) \sin^4(Jt)]  - 4 \sin^6(Jt) \cos^2(Jt)  \log[4 \sin^6(Jt)  \cos^2(Jt)].  \end{align}

This is shown in dotted curve of fig.1(d). We begin with a separable input
state(Entropy of entanglement(S) =0) after which the value of 'S' increases and attains a maximum value of 2.03064 at $Jt=0.785212$, 
it then decreases and eventually becomes equal to zero at $Jt=1.57061$. Thus the
state becomes disentangled at this point of time. At later times we see a periodic behavior and
the system gets entangled and disentangled periodically.

 Now we consider the logarithmic negativity for each of the four photon states \\
 $\{\psi_{in}\}=\{|2,2>, |3,1>, |1,3>, |4,0>, |0,4>,$ and $(|4,0>+|0,4>)/\sqrt{2}$(four photon N00N state)$\}$.

{\bf Case-2(a.)} For the  input state as $|\psi_{in}> = |2,2>$, the possible output states are,
 
 $|\psi_{out}> = a_1 |0,4> + a_2|3,1> + a_3 |2,2> +a_4 |1,3> +a_5 |4,0>.$
where, $ a_1 = -\sqrt{6} \sin^2(Jt) \cos^2(Jt), \\ with, 
a_2 = i\sqrt(6)(\sin^3(Jt) \cos(Jt)-\sin(Jt)\cos^3(Jt)), a_3 = \cos^4(Jt)+ \sin^4(Jt)-4\sin^2(Jt)\cos^2(Jt), \\
a_4=i\sqrt(6)(\sin^3(Jt) \cos(Jt)-\sin(Jt)\cos^3(Jt)),$ and $a_5 =  -\sqrt{6} \sin^2(Jt) \cos^2(Jt).$ 

Then the log negativity entanglement of this system , \vspace{-0.7cm}
\be E_N  = \log_2[1+2(a_1a_2+a_1a_3+a_1a_4+a_1a_5+a_2a_3+a_2a_4+a_2a_5+a_3a_4+a_3a_5+a_4a_5] \ee
 is shown in thick curve of Fig.1(b). The system starts at  $t = 0$, with a separable input
state and log negativity zero, which increases at  $Jt = 0.200242$ to  $E_N = 1.15181$ and for $Jt =0.325477$  dips at $E_N = 1$ due to Hong-Ou-Mandel
interference\cite{Hong}. The $E_N$ increases with time and attains a value of 1.77519 for $Jt = 0.601094$, this is the maximally entangled state, and for $Jt = 0.785212$ 
again we get dips at $E_N=1.7277$ due to Hong-Ou-Mandel interference. At later times,
we see a periodic behaviorr
, attributed to the inter-waveguide coupling(J). Because of involvement of four photons, we can see the double the interference effect of the  two photon system.

{\bf Case-2(b.)} For the input state as $|\psi_{in}> = |3,1>$, the possible output states are

$|\psi_{out}> =  b_1 |0,4> + b_2|3,1> + b_3 |2,2> +b_4 |1,3> +b_5 |4,0>.$
where, $ b_1 =2i\sin^3(Jt) \cos(Jt),\\
b_2 = \sin^4(Jt)-3\sin^2(Jt)\cos^2(Jt),  b_3 = i\sqrt(6)(\sin^3(Jt) \cos(Jt)-\sin(Jt)\cos^3(Jt)), \\
b_4=\cos^4(Jt)-3\sin^2(Jt)\cos^2(Jt),$  and $b_5 =  -2i\sin(Jt) \cos^3(Jt).$

The Entanglement for this system is, 
\be E_N = \log_2[1+2(b_1b_2+b_1b_3+b_1b_4+b_1b_5+b_2b_3+b_2b_4+b_2b_5+b_3b_4+b_3b_5+b_4b_5)] \ee

 and  is shown in dotted curve of Fig.1(b). We see the small difference in interference pattern with the  $|2,2>$ state. The  Hong-Ou-Mandel effect
 and the maximally entangled state occur at lower values.
 
 \vspace{0.3cm}

{\bf Case-2(c.)} Now we take the input state as $|\psi_{in}> = |4,0>$, theoutput state is

$|\psi_{out}>=c_1|0,4>+c_2|3,1>+c_3|2,2>+c_4|1,3>+c_5|4,0>.$
here, $c_1=\sin^4(Jt),\\
c_2=2i\sin^3(Jt)\cos(Jt), c_3 = -\sqrt{6} \sin^2(Jt) \cos^2(Jt), c_4=-2i\sin(Jt)\cos^3(Jt),$ and $c_5 = \cos^4(Jt).$

The Entanglement for this system 
\be E_N = \log_2[1+2(c_1c_2+c_1c_3+c_1c_4+c_1c_5+c_2c_3+c_2c_4+c_2c_5+c_3c_4+c_3c_5+c_4c_5)] \ee
 is shown in thin curve of Fig.1(b).  
 
 {\bf Case-2(d.)} For four photon input NOON state:

$|\psi_{in}> = (|4,0> + |0,4>)/\sqrt{2*4!},$  the output state is, 
$|\psi_{out}>=d_1|0,4>+d_2|3,1>+d_3|2,2>+d_4|1,3>+d_5|4,0>$
here, $d_1=(\sin^4(Jt)+\cos^4(Jt))/\sqrt{2},\\ 
d_2 = i\sqrt(2)(\sin^3(Jt) \cos(Jt)-\sin(Jt)\cos^3(Jt)), d_3 = -\sqrt{3}\sin^2(Jt)\cos^2(Jt), \\
d_4=i\sqrt(2)(\sin^3(Jt) \cos(Jt)-\sin(Jt)\cos^3(Jt)),$ and $d_5 =(\sin^4(Jt) + \cos^4(Jt))/\sqrt{2}.$

The logarithmic negativity for this system, 
\be E_N = \log_2[1+2(d_1d_2+d_1d_3+d_1d_4+d_1d_5+d_2d_3+d_2d_4+d_2d_5+d_3d_4+d_3d_5+d_4d_5)] \ee
 is shown in dotdashed curve of Fig.1(c). In this case , unlike for the two photon NOON state, the entanglement never goes to zero. 
This means that increasing the number of photons in a NOON state gives a more robust entanglement which is sustained at large times. 
To show this we calculate the entropy for entanglement  and Logarithmic negativity for a 3-photon and a five photon NOON state.
The entropy of entanglement for three photon system 
is shown in thin curve of fig.1(d).
 For three photon input NOON state the entanglement  is, \vspace{-0.7cm}

\be E_N = \log_2[1+2(b_1b_2+b_1b_3+b_1b_4+b_2b_3+b_2b_4+b_3b_4)], \ee where,
 $b_1=(\cos^3(Jt)+i\sin^3(Jt))/\sqrt{2},\\ 
b_2 = -i\sqrt(\frac{3}{2})(\sin^2(Jt) \cos(Jt)+i\sin(Jt)\cos^2(Jt)), b_3 = -i\sqrt(\frac{3}{2})(\sin^2(Jt))$
 and is shown in dotted curve of Fig.1(c). 
 Similarly for a  five photon system as an input(i.e.,$N=5$):
 he entropy of entanglement for five photon system is, shown in thick curve of fig.1(d).

 For five photon input NOON state, the entanglement for this system is,
\be E_N = \log_2[1+2(e_1e_2+e_1e_3+e_1e_4+e_1e_5+e_1e_6+e_2e_3+e_2e_4+e_2e_5+e_2e_6+e_3e_4+e_3e_5+e_3e_6+e_4e_5+e_4e_6+e_5e_6)] \ee
 $e_1=(\cos^5(Jt)+i\sin^5(Jt))/\sqrt{2}, 
e_2 =i\sqrt(\frac{5}{2})(\sin^4(Jt) \cos(Jt)-i\sin(Jt)\cos^4(Jt)),\\ e_3 = -\sqrt(5)(\sin^2(Jt) \cos^3(Jt)-i\sin^3(Jt)\cos^2(Jt)),
e_4 = -\sqrt(5)(\sin^2(Jt) \cos^3(Jt)-i\sin^3(Jt)\cos^2(Jt)),\\ e_5 =i\sqrt(\frac{5}{2})(\sin^4(Jt) \cos(Jt)-i\sin(Jt)\cos^4(Jt)),$ 
 and $e_6 =(\cos^5(Jt) -i \sin^5(Jt))/\sqrt{2}.$  The Logarithmic entropy  is shown in thin curve of Fig.1(c). 
 We see that as the photon number increases the NOON state gets more and more robust and shows that "high-noon" states can be used for more precision measurements. 
 These results are relevant in light of the recent experimental detection of entangled 5-photon "NOON" states\cite{NOON}.  

\subsection{Entanglement properties for Input Thermal States} 
Now we consider the initial state $|\rho(0)>$ to be the two mode thermal state. Then, in  the TFD formalism the time evolved state $\rho(t)$ is
\vspace{-0.3cm}
\begin{align} \nonumber |\rho(t)> & = e^{-iG(\theta)} e^{-iHt} \otimes e^{-i\tilde{H}t} |\rho(0)>
 \\&  = \sum^N_{n_a=0} \frac{\bar{n}^{n_a}_a}{(\bar{n}_a+1)^{n_a+1}} \sum^N_{n_b=0} \frac{\bar{n}^{n_b}_b}{(\bar{n}_b+1)^{n_b+1}}
e^{-iHt} |n_a,n_b> <n_a,n_b| e^{iHt}  
 \\&  =  \sum^N_{n_a=0} C_{n_a,n_a} \frac{\bar{n}^{n_a}_a}{(\bar{n}_a+1)^{n_a+1}} \sum^N_{n_b=0} C_{n_b,n_b} \frac{\bar{n}^{n_b}_b}{(\bar{n}_b+1)^{n_b+1}}
  |n_a,N-n_a><n_b,N-n_b| 
\end{align}

where, $G(\theta)=-i \theta(\tilde{a}a - \tilde{a}^\dag a^\dag + \tilde{b}b - \tilde{b}^\dag b^\dag),$ and $\bar{n}_a$ and $\bar{n}_b$
are thermal distribution functions .
$ C_{n_a,n_a} = \frac{\xi^{n_a}}{(1+|\xi|^2)^{N/2}} \left(
\begin{array}{rcl}
N \\
 n_a \\
\end{array}
\right)^{\frac{1}{2}};$ \hspace{0.7cm} and \hspace{0.4cm}  $N=n_a+n_b.$

The entanglement properties are calculated by taking the partial transpose of $\rho$ and finding  its eigenvalues, which are given by
\be \lambda_{n_an_a} =  \frac{(\bar{n}_a)^{2n_a}}{(\bar{n}_a+1)^{2(n_a+1)}}  \{ \frac{N!}{(N-n_a)! n_a!} \sin^{2n_a}(Jt) \cos^{2N-2n_a}(Jt)\};  (for,  n_a=n_b) \ee
\small{\be
 \lambda_{n_an_b} = \pm [\frac{\bar{n}^{n_a}_a}{(\bar{n}_a+1)^{n_a+1}} \frac{\bar{n}^{n_b}_b}{(\bar{n}_b+1)^{n_b+1}}
 \{\frac{N!}{\sqrt{(N-n_a)! (N-n_b)! n_a! n_b! }} \sin^{n_b+n_a}(Jt) \cos^{2N-n_b-n_a}(Jt)\}];(n_a\ne n_b) \ee} \vspace{0.3cm}
 The  corresponding von-Neumann entropy  is:
\vspace{-0.3cm}
 \begin{align} \nonumber S &=-\sum^N_{i=0} \lambda_i \log \lambda_i = -\sum^N_{n_a = 0} \frac{(\bar{n}_a)^{2n_a}}{(\bar{n}_a+1)^{2(n_a+1)}}
 [\frac{N!}{(N-n_a)! n_a!} \sin^{2n_a}(Jt) \cos^{2N-2n_a}(Jt)] 
 \\&  \hspace{1.2cm} \times \{ \log[\frac{(\bar{n}_a)^{2n_a}}{(\bar{n}_a+1)^{2(n_a+1)}} [\frac{N!}{(N-n_a)! n_a!} \sin^{2n_a}(Jt) \cos^{2N-2n_a}(Jt)]] \} .\label{entropy}
\end{align}

\begin{figure}
\begin{center}
\includegraphics[width=3in,height=2.2in]{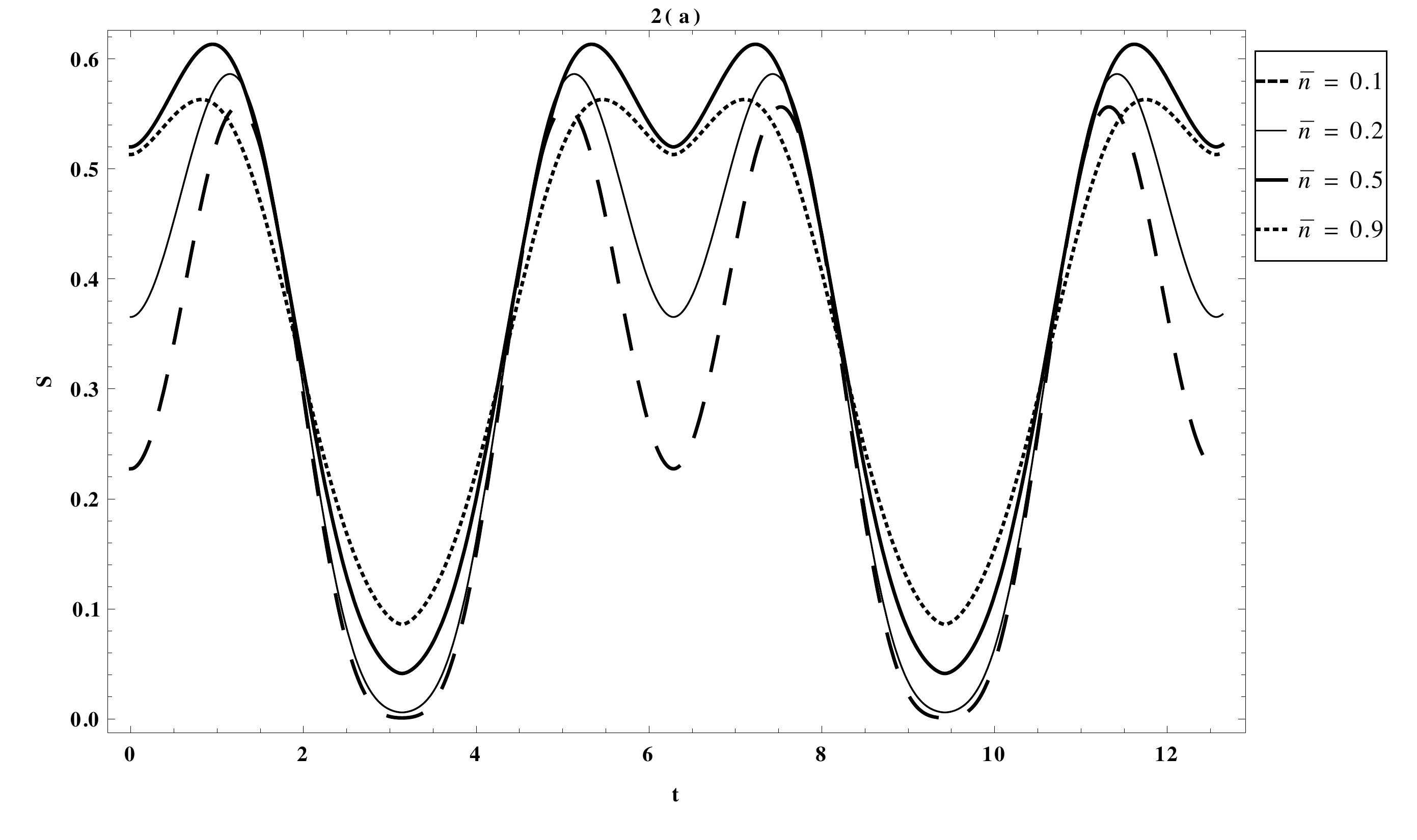} \hspace{0.2cm} \includegraphics[width=3in,height=2.2in]{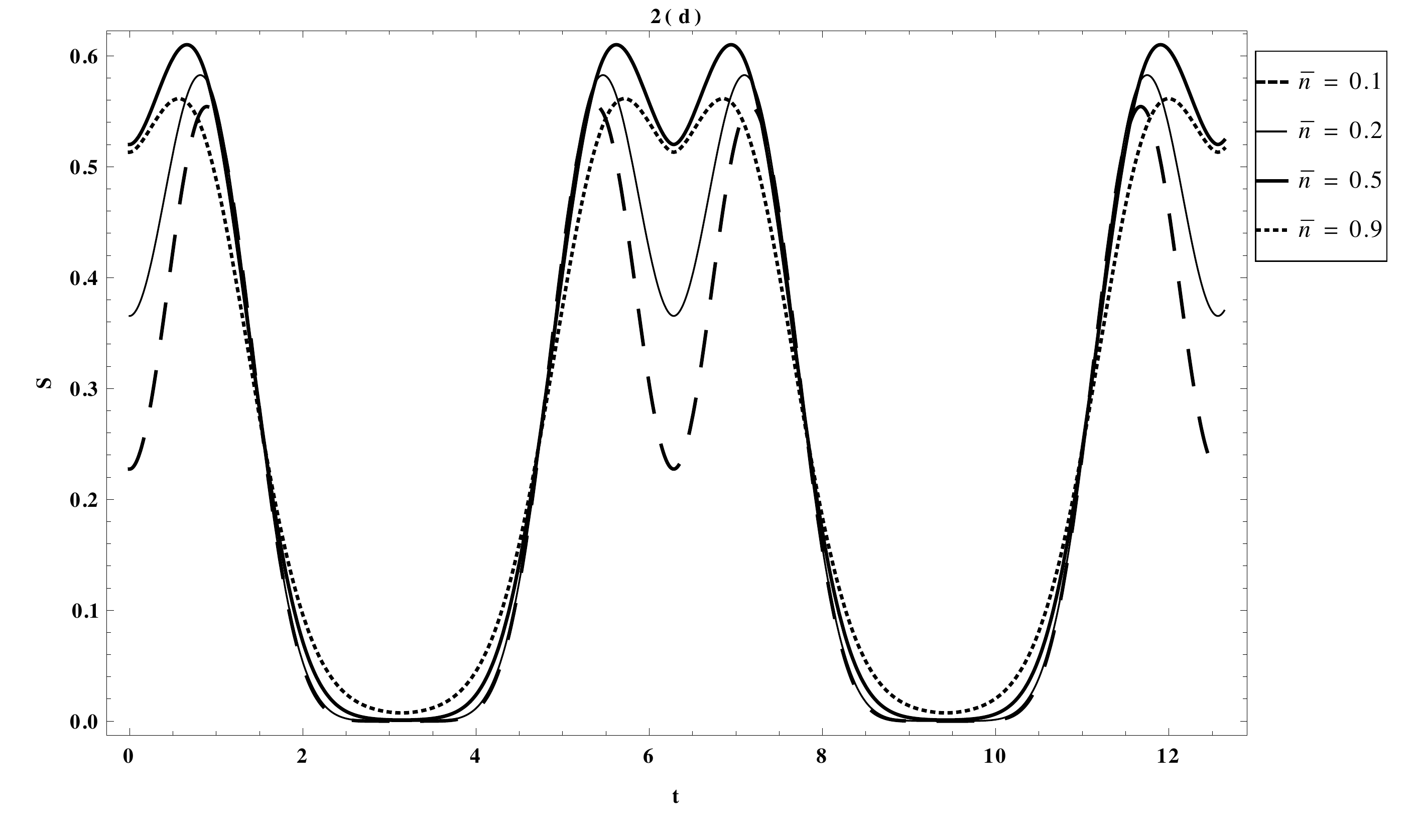} \\ \vspace{1cm}
\includegraphics[width=3in,height=2.2in]{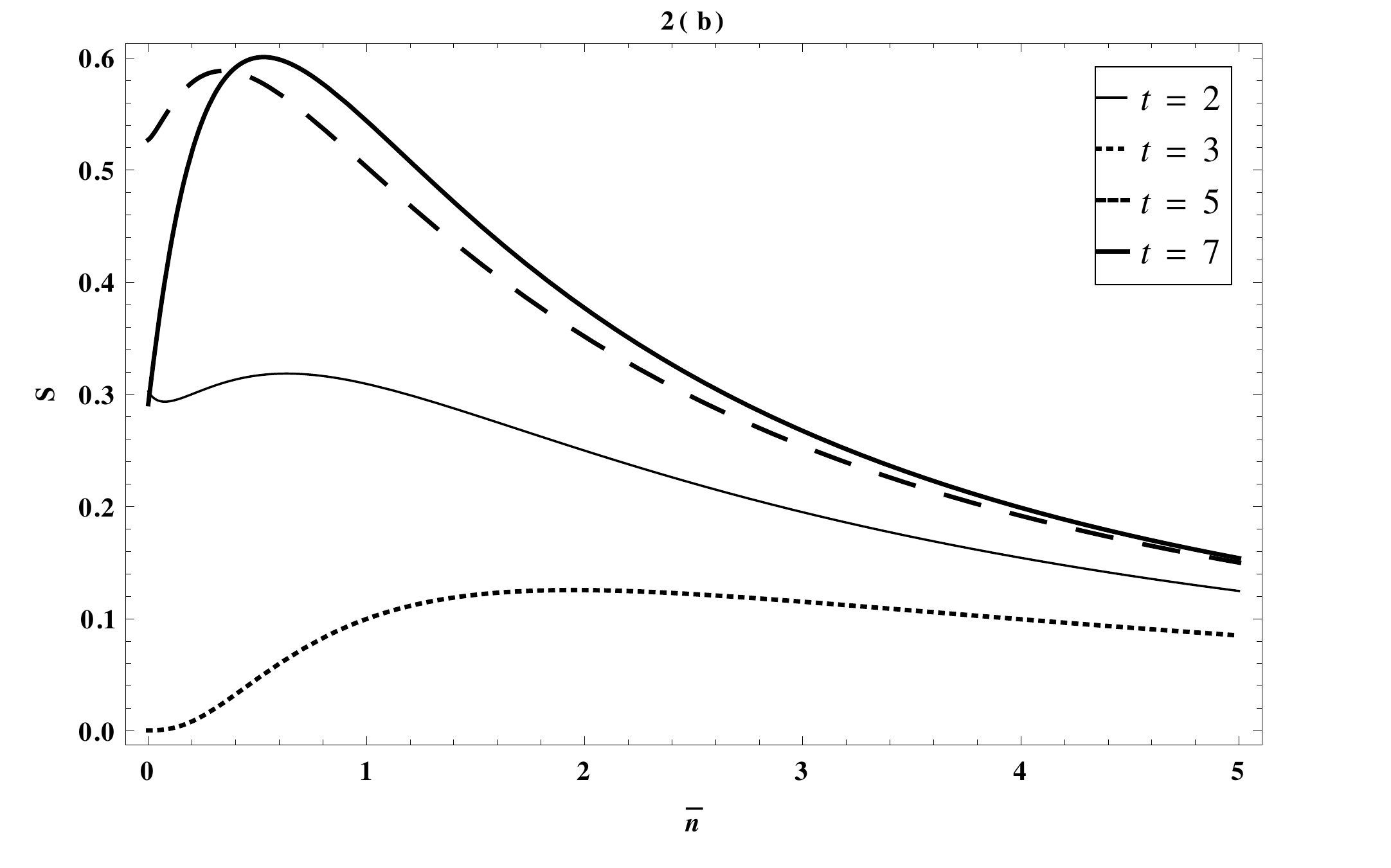}  \hspace{0.2cm} \includegraphics[width=3in,height=2.2in]{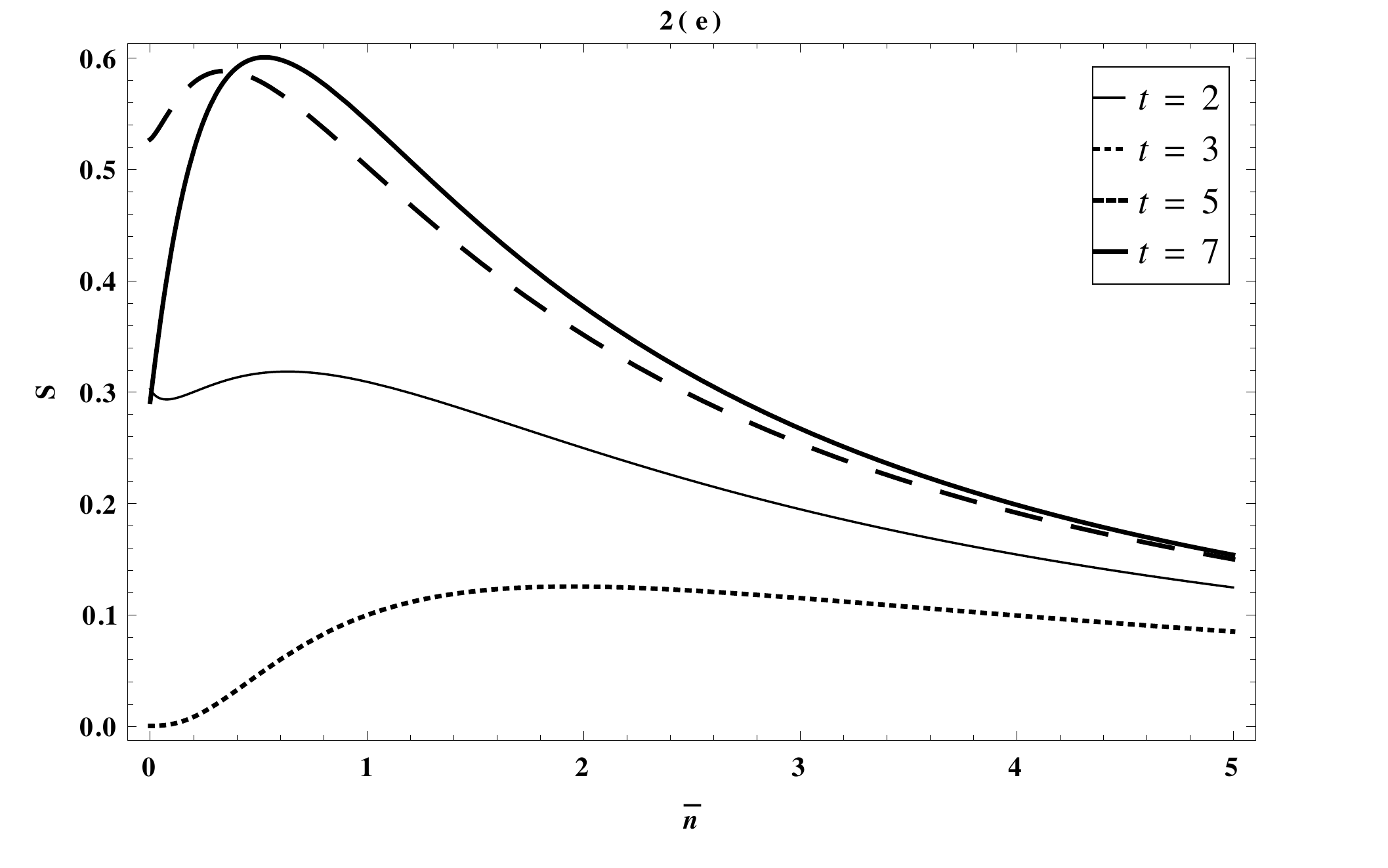} \\ \vspace{1cm}
\includegraphics[width=3in,height=2.2in]{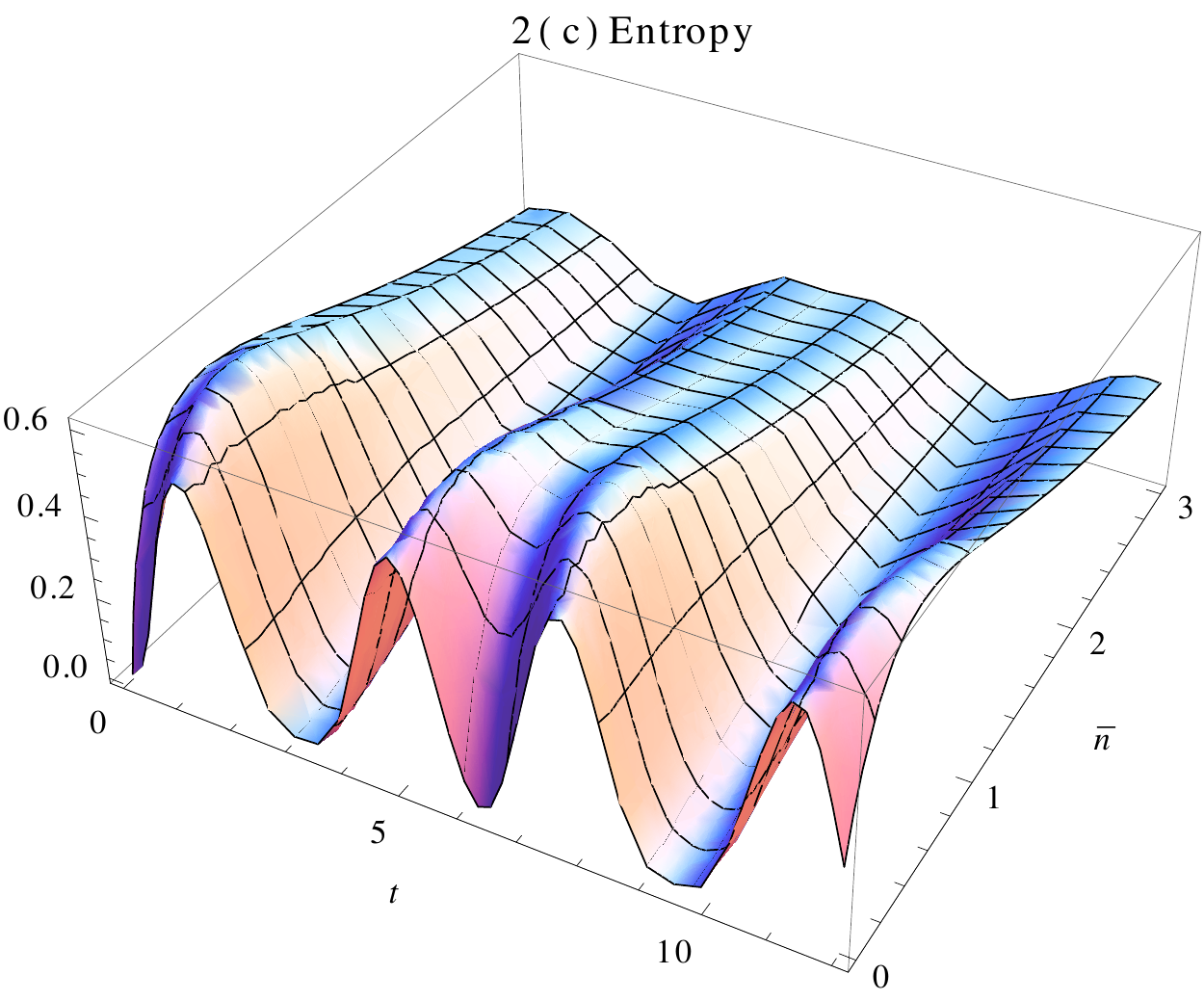}  \hspace{0.2cm} \includegraphics[width=3in,height=2.2in]{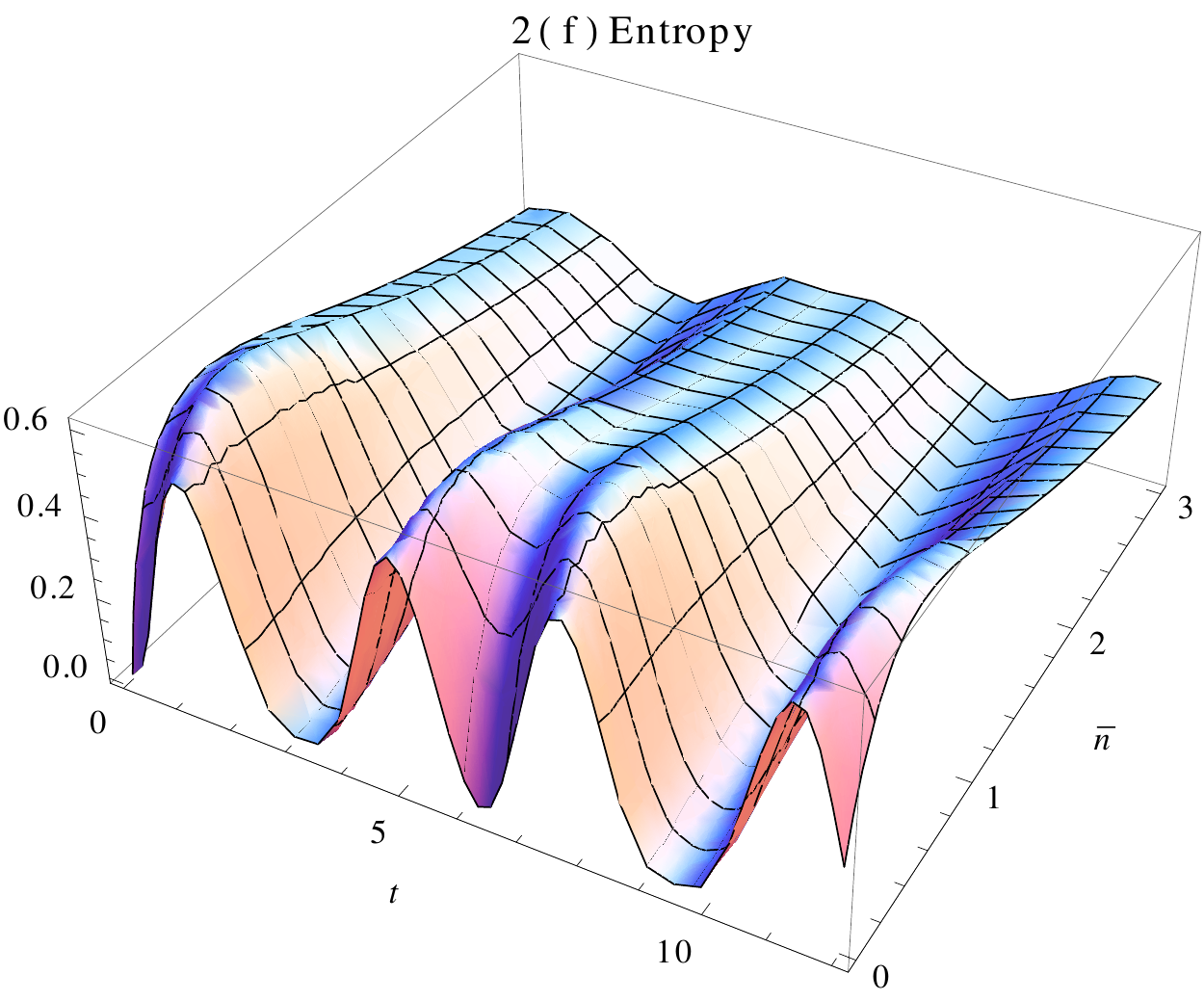} \\
\end{center}
\vspace{-0.5cm}
\tiny{\caption{For input two photon system: fig(a) shows the time evolution of entropy of entanglement(S) for different values of $\bar{n}$,
fig(b) shows entropy of entanglement(S) vs. Average number($\bar{n}$) for different values of t,
fig(c) shows 3D-plot of entropy of entanglement in function of $\bar{n}$ and t; 
For input four photon system: fig(d.) Time evolution of entropy of entanglement(S) for different values of $\bar{n}$,
fig(e) Entropy of entanglement(S) vs. Average number($\bar{n}$) for different values of t, 
fig(f) shows 3D-plot of entropy of entanglement in function of $\bar{n}$ and t.}} \label{2} 
\end{figure}

{\bf Case-3(a.)} {\bf Two photon system input(i.e., $N=2$):}
  
 The entropy of entanglement for the two photon  thermal input state  is  \vspace{-0.3cm}
\begin{align} \nonumber S  = & - \frac{1}{(\bar{n}+1)^2} \cos^4(Jt) \log[\frac{1}{(\bar{n}+1)^2} \cos^4(Jt)]
-\frac{2\bar{n}^2}{(\bar{n}+1)^4} \sin^2(Jt) \cos^2(Jt)  \log[\frac{2\bar{n}^2}{(\bar{n}+1)^4} \sin^2(Jt) \cos^2(Jt)] 
\\& - \frac{1}{(\bar{n}+1)^2} \sin^4(Jt) \log[\frac{1}{(\bar{n}+1)^2} \sin^4(Jt)] \end{align}

 and  is shown in Figs.2(a,b,c). 
 \vspace{0.3cm}
 
{\bf Case-3(b.)} {\bf  Four photon input(i.e., $N=4$):}

The entropy of entanglement for four photon thermal input  is  \vspace{-0.3cm}

\begin{align} \nonumber  S  = &-\frac{1}{(\bar{n}+1)^2} \cos^8(Jt)\log[\frac{1}{(\bar{n}+1)^2} \cos^8(Jt)] 
- \frac{\bar{n}^8}{(\bar{n}+1)^{10}} \sin^8(Jt)  \log[\frac{\bar{n}^8}{(\bar{n}+1)^{10}} \sin^8(Jt) ] 
\\& \nonumber - \frac{4\bar{n}^2}{(\bar{n}+1)^{4}} \sin^2(Jt) \cos^6(Jt)  \log[\frac{4\bar{n}^2}{(\bar{n}+1)^{4}} \sin^2(Jt) \cos^6(Jt) ] 
\\& \nonumber - \frac{6\bar{n}^4}{(\bar{n}+1)^{6}}  \sin^4(Jt) \cos^4(Jt)  \log[\frac{6\bar{n}^4}{(\bar{n}+1)^{6}}  \sin^4(Jt) \cos^4(Jt) ]
\\&  - \frac{4\bar{n}^6}{(\bar{n}+1)^{8}}  \sin^6(Jt) \cos^2(Jt)  \log[\frac{4\bar{n}^6}{(\bar{n}+1)^{8}}  \sin^6(Jt) \cos^2(Jt)]   \end{align}

and is shown in Figs.2(d,e,f).

 These figures show that for low values of $\bar{n_a}$ and $\bar{n_b}$, the system sustains entanglement, but as the system gets more thermalized,
 it decoheres and the entropy of entanglement tends to zero. The thermal effect mimics a damping effect in the system.

\subsection{Two coupled waveguides  with damping($\gamma \ne 0$)}
We consider now, losses in the coupled waveguides due to system-reservoir interaction with '$\gamma$' as the rate of loss due to
the material of the waveguide.
The time evolution of the density operator equation is

  \be |\dot{\rho}> = -i \hat{H^\prime}|\rho> \label{eqnofmtn}\ee
  with , \vspace{-.5cm}  \begin{align} \nonumber  \hat{H^\prime} = & \omega (a^\dag a +b^\dag b - \tilde{a}^\dag \tilde{a}  - \tilde{b}^\dag \tilde{b}) + J(a^\dag b + b^\dag a  - \tilde{a}^\dag \tilde{b} - \tilde{b}^\dag \tilde{a})
  \\& - i\gamma (a^\dag a - a \tilde{a} +a^\dag \tilde{a}^\dag + \tilde{a}^\dag \tilde{a} + b^\dag b - b \tilde{b} +b^\dag \tilde{b}^\dag +  \tilde{b}^\dag \tilde{b} )  \label{hamiltonian} \end{align}

the following transformations,  \vspace{-0.2cm}
 
 \be a=\frac{A+B}{\sqrt{2}},\hspace{.5cm} a^\dag=\frac{A^\dag+B^\dag}{\sqrt{2}},\hspace{.5cm}
 \tilde{a} = \frac{\tilde{A}+\tilde{B}}{\sqrt{2}}, \hspace{.5cm} \tilde{a}^\dag = \frac{\tilde{A}^\dag+\tilde{B}^\dag}{\sqrt{2}} \label{rot1} \ee
 \vspace{-0.1cm}
\be  b=\frac{-A+B}{\sqrt{2}}, \hspace{.5cm} b^\dag=\frac{-A^\dag+B^\dag}{\sqrt{2}}, \hspace{.5cm}
\tilde{b} = \frac{-\tilde{A}+\tilde{B}}{\sqrt{2}},  \hspace{.5cm}  \tilde{b}^\dag = \frac{-\tilde{A}^\dag+\tilde{B}^\dag}{\sqrt{2}} \label{rot2}, \ee
give the Hamiltonian   \hspace{-0.3cm} \begin{align} \nonumber  \hat{H^{\prime\prime}} = & \omega (A^\dag A +B^\dag B - \tilde{A}^\dag \tilde{A} -\tilde{B}^\dag  \tilde{B}) + J(-A^\dag A + B^\dag B +  \tilde{A}^\dag \tilde{A} - \tilde{B}^\dag \tilde{B} )
  \\& -i \gamma (A^\dag A + B^\dag B  +  \tilde{A}^\dag \tilde{A} + \tilde{B}^\dag \tilde{B}  - A \tilde{A}- B \tilde{B} +A^\dag \tilde{A}^\dag  +B^\dag \tilde{B}^\dag)  \label{hamiltonian1} \end{align}
  
  We diagonalise this Hamiltonian by applying a squeezing (Bogolubov) transformation mixing the real and tilde fields.   \vspace{-0.2cm}
  \be  D = \mu_1 A + \nu^*_1 \tilde{A}^\dag,\hspace{.5cm}   D^\dag = \mu^*_1 A^\dag + \nu_1 \tilde{A}, \hspace{.5cm}
  \tilde{D} = \mu_1 \tilde{A} + \nu^*_1 A^\dag \hspace{.2cm}  \& \hspace{.2cm}  \tilde{D}^\dag = \mu^*_1 \tilde{A}^\dag + \nu_1 A \label{munu1}\ee
   \be E = \mu_2 B + \nu^*_2 \tilde{B}^\dag, \hspace{.5cm}  E^\dag = \mu^*_2 B^\dag + \nu_2 \tilde{B}, \hspace{.5cm}
   \tilde{E} = \mu_2 \tilde{B} + \nu^*_2 B^\dag \hspace{.2cm} \&  \hspace{.2cm}  \tilde{E}^\dag = \mu^*_2 \tilde{B}^\dag + \nu_2 B \label{munu2},\ee
   where, \vspace{-0.2cm}  
 \be \mu_1 = cosh r_1 = \sqrt{\frac{|(\omega -J-i\gamma|^2}{|-i\gamma-\sqrt{\gamma^2+(\omega-J)^2}|^2}}, \hspace{.7cm}
 \nu_1 = sinh r_1 = \sqrt{\frac{-|i\gamma|^2}{|-i\gamma+\sqrt{\gamma^2+(\omega-J)^2}|^2}} \label{MN1} \ee
 \vspace{-0.1cm} 
   \be \mu_2 = cosh r_2 = \sqrt{\frac{|(\omega +J-i\gamma|^2}{|-i\gamma-\sqrt{\gamma^2+(\omega+J)^2}|^2}}, \hspace{.7cm}
   \nu_2 = sinh r_2 = \sqrt{\frac{-|i\gamma|^2}{|-i\gamma+\sqrt{\gamma^2+(\omega+J)^2}|^2}}\label{MN2}  \ee
   and $r_1 $ and $r_2$ are the squeezing parameters and  $|\mu_1|^2 -|\nu_1|^2 =1 $ and $|\mu_2|^2 -|\nu_2|^2 =1.$ The final Hamiltonian is written as,
   \vspace{-0.3cm} 
   \begin{align} \nonumber \hat{H}_f & = S^{-1}(r_1) H_A S(r_1) +  S^{-1}(r_2) H_B S(r_2) 
   \\& =\Omega^2_1 (D^\dag D) + \Omega^2_2 (\tilde{D}^\dag \tilde{D}) +\Omega^2_3 ((E^\dag E) + \Omega^2_4 (\tilde{E}^\dag \tilde{E}) \label{hamiltonianfinal} .\end{align}
   
where, \be S(r_1) = exp[r_1 K_{A_+}-r^*_1 K_{A_-}] = Exp[r_1A^\dag \tilde{A}^\dag - r^*_1 A \tilde{A}]; \ee
 \be S(r_2) = exp[r_2 K_{B_+}-r^*_2 K_{B_-}] = Exp[r_2B^\dag \tilde{B}^\dag - r^*_2 B \tilde{B}]; \ee
 
 and $\Omega_1 = - \frac{\sqrt{\gamma^2 + (\omega - J)^2}}{2} -\frac{i\gamma}{2}$, 
 $\Omega_2 = \frac{\sqrt{\gamma^2 + (\omega - J)^2}}{2} -\frac{i\gamma}{2}$, $\Omega_3 = - \frac{\sqrt{\gamma^2 + (\omega + J)^2}}{2} -\frac{i\gamma}{2}$,
 $\Omega_4 = \frac{\sqrt{\gamma^2 + (\omega + J)^2}}{2} -\frac{i\gamma}{2}$; 
 The generators of the SU(1,1) algebra  in terms of the modes $A$ and $B$ are given by 

\be K_{A-} = A\tilde{A}, \hspace{.7cm} K_{A+} = A^\dag \tilde{A}^\dag, \hspace{.7cm}  K_{A3} = \frac{A^\dag A  + \tilde{A}^\dag \tilde{A} +1}{2} \ee
 \vspace{-0.2cm} 
\be K_{B-} = B\tilde{B}, \hspace{.7cm} K_{B+} = B^\dag \tilde{B}^\dag, \hspace{.7cm} K_{B3} = \frac{B^\dag B  + \tilde{B}^\dag \tilde{B} +1}{2}, \ee 
and  satisfy the commutation relations 
 \vspace{-0.4cm}
\be [K_{A-},K_{A+}] = 2K_{A3}, \hspace{.2cm} [K_{A3},K_{A\pm}] =\pm K_{A\pm} ; \hspace{.3cm}  [K_{B-},K_{B+}] = 2K_{B3}, \hspace{.2cm} [K_{B3},K_{B\pm}] =\pm K_{B\pm}. \ee 

The Casimir operators are   \vspace{-0.4cm} \be K_{Ao} = (A^\dag A  - \tilde{A}^\dag \tilde{A}), \hspace{.7cm}  K_{Bo} = (B^\dag B  - \tilde{B}^\dag \tilde{B}). \ee
Then the solution of eq.(\ref{eqnofmtn}) becomes \vspace{-0.4cm}
\be |\rho(t)> = K(t)e^{[ \eta_{A3} K_{A3}  + \eta_{A-} K_{A-}+ \eta_{A+} K_{A+} + \eta_{B3} K_{B3} + \eta_{B-} K_{B-} + \eta_{B+} K_{B+} ]}|\rho(0)>
 \label{eqnofmtnsu11}
\ee
where, \vspace{-1.2cm}
\begin{align} \nonumber  & K(t) = e^{-i\omega t(K_{A0} +K_{B0}) +2iJt(N_A+\tilde{N}_B-1)}, \hspace{0.3cm} \eta_{A-} =\gamma t = \eta_{B-}, \\& 
 \eta_{A3} = -2(\gamma +iJ)t =  \eta_{B3}, \hspace{0.4cm} \eta_{A+} = -\gamma t = \eta_{B+}. \label{prmts} \end{align}
 
 By using the $SU(1,1)$  disentanglement formula\cite{Perelomov}, one can write eq.(\ref{eqnofmtnsu11}) as, \vspace{-0.5cm}
 
\begin{align} \nonumber |\rho(t)> = & \{K(t) exp[\Gamma_{A+} K_{A+}] exp[\ln(\Gamma_{A3}) K_{A3}] 
exp[\Gamma_{A-} K_{A-}] \\& \otimes e^[\Gamma_{B+} K_{B+}] exp[\ln(\Gamma_{B3} K_{B3})] exp[\Gamma_{B-} K_{B-}] \}|\rho(0)> \label{eqnofmtnsu11disentgl}\end{align} 
here,  \vspace{-0.5cm}

\be \Gamma_{i\pm} = \frac{2\eta_{i\pm} \sinh\phi_i}{2\phi_i \cosh\phi_i-\eta_{i3} \sinh\phi_i}  \hspace{.7cm} \& \hspace{.7cm}
 \Gamma_{i3} = \left( \frac{{2\phi_i}}{{2\phi_i} \cosh\phi_i - \eta_{i3}\sinh\phi_i} \right)^2 \label{pmts} \ee
 
with  \vspace{-0.2cm} 
\be \phi_i^2 = \frac{\eta_{i3}^2}{4}-\eta_{i+}\eta_{i-},\label{qnt} \ee subscript  i labels A, B.

We consider an  initial state $ |\rho(0) > = \sum_{m,n}^N \rho_{m,n}(0) |m,\tilde{m},n,\tilde{n}> $, in TFD notation.
This gives us an exact solution of the density matrix  :

\begin{align} \nonumber \rho_{m,n}(t) = &  C(t)\sum_{q^\prime = 0}^{min(m^\prime,n^\prime)} \sum_{p^\prime = 0}^\infty [
 \left(
\begin{array}{rcl}
m^\prime +p^\prime - q^\prime\\
    p^\prime      \\
\end{array}
\right)
\left(
\begin{array}{rcl}
n^\prime +p^\prime - q^\prime\\
p^\prime\\
\end{array}
\right)
 \left(
\begin{array}{rcl}
m^\prime\\
q^\prime\\
\end{array}
\right)
\left(
\begin{array}{rcl}
n^\prime\\
q^\prime\\
\end{array}
\right) ]^{\frac{1}{2}}
\\& \nonumber \times \sum_{q = 0}^{min(m,n)} \sum_{p = 0}^\infty [
 \left(
\begin{array}{rcl}
m +p - q\\
p\\
\end{array}
\right)
\left(
\begin{array}{rcl}
n +p - q\\
p\\
\end{array}
\right)
 \left(
\begin{array}{rcl}
m\\
q\\
\end{array}
\right)
\left(
\begin{array}{rcl}
n\\
q\\
\end{array}
\right) ]^{\frac{1}{2}}
\\& \nonumber \times [\Gamma_{A+}]^{p^\prime} [\Gamma_{A3}]^\frac{(m^\prime+n^\prime-2q^\prime+1)}{2} [\Gamma_{A-}]^{q^\prime}
[\Gamma_{B+}]^p [\Gamma_{B3}]^\frac{(m+n-2q+1)}{2} [\Gamma_{B-}]^q \\& \times \rho_{m+p-q, m^\prime+p^\prime-q^\prime, n+p-q, n^\prime+p^\prime-q^\prime} (0)
\end{align}

where $C(t)$is an overall phase factor due to $K(t) = e^{-i\omega t(K_{A0} +K_{B0}) +2iJt(N_A+\tilde{N}_B-1)}.$ This is the exact solution for the density matrix of the coupled lossy system of waveguides.
\subsubsection{Calculation of Entanglement of the System for $\gamma\ne0$} 

The entanglement properties are calculated by first  taking trace of $\rho(t)$ over the tilde space and then taking the partial transpose of $\rho$ 
and calculating the eigenvalues of the resulting matrix similar to the case without damping. The eigenvalues are 
\be \lambda_{m_am_a} =  \frac{N!}{(N-m_a)! m_a!} \sinh^{2m_a}(\theta) \cosh^{2N-2m_a}(\theta);  (for,  m_a=n_b) \;and 
\; \; \; \theta = \left(\sqrt{2}\gamma +iJ \right) t.\ee \be
 \lambda_{m_an_b} = \pm \frac{N!}{\sqrt{(N-m_a)! (N-n_b)! m_a! n_b! }} \sinh^{m_a+n_b}(\theta) \cosh^{2N-m_a-n_b}(\theta); (for,  m_a\ne n_b) \ee
  The entropy of entanglement of the  the system is \vspace{-0.3cm}
 \begin{align} \nonumber S &=-\sum^N_{i=0} \lambda_i \log \lambda_i =  -\sum^N_{m_a = 0} [\frac{N!}{(N-m_a)! m_a!} \sinh^{2m_a}(\theta) \cosh^{2N-2m_a}(\theta)] 
 \\&  \hspace{1.2cm} \times \{ \log[\frac{N!}{(N-m_a)! m_a!}] +2m_a \log[\sinh\theta] + (2N-2m_a)\log[\cosh\theta] \} \label{entropy}
\end{align}
 
 Thus, for two photon state as an input state, the entropy of entanglement of the system is  \vspace{-0.5cm}

\be S  =- 4 \cosh^4(\theta) \log[\cosh(\theta)] - 2 \sinh^2(\theta) \cosh^2(\theta)  \log[2\cosh^2(\theta) \sinh^2(\theta)] - 4 \sinh^4(\theta)  \log[\sinh(\theta)] \ee

and is shown in thin curves of Figs.3(a,b,c,d).
\begin{center}
\begin{figure}
\includegraphics[width=2.7in,height=2in]{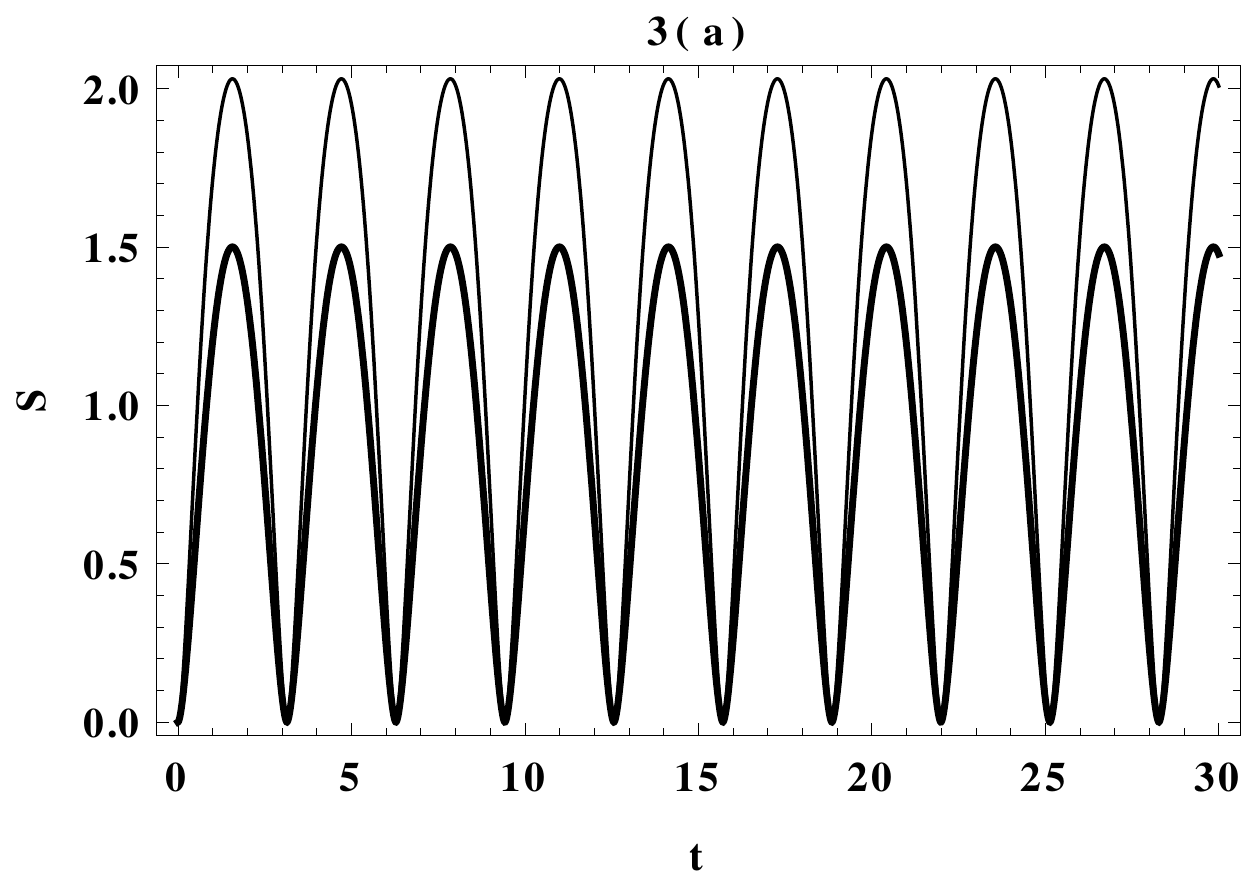} \hspace{0.5cm}  \includegraphics[width=3.3in,height=2in]{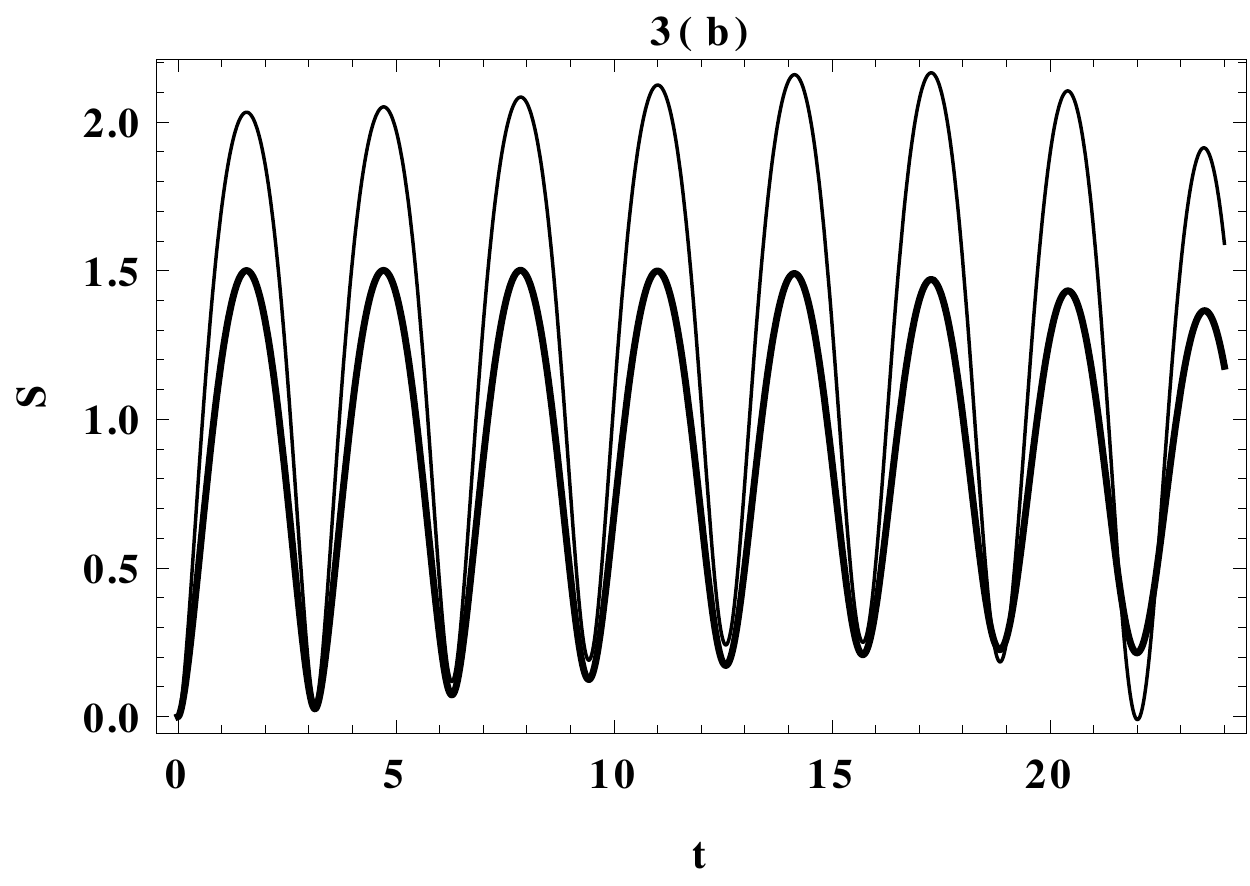} \\
\includegraphics[width=2.7in,height=2in]{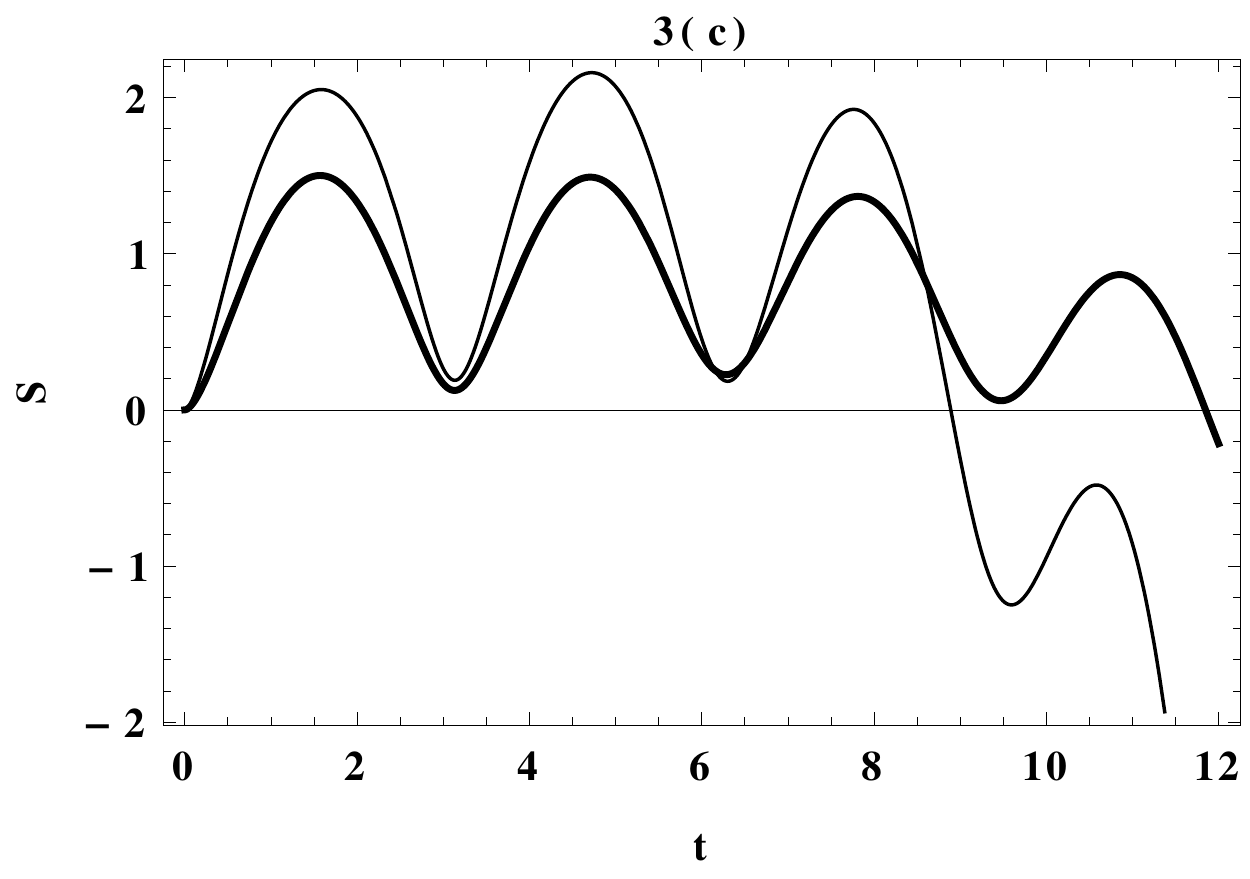} \hspace{0.5cm}  \includegraphics[width=3.3in,height=2in]{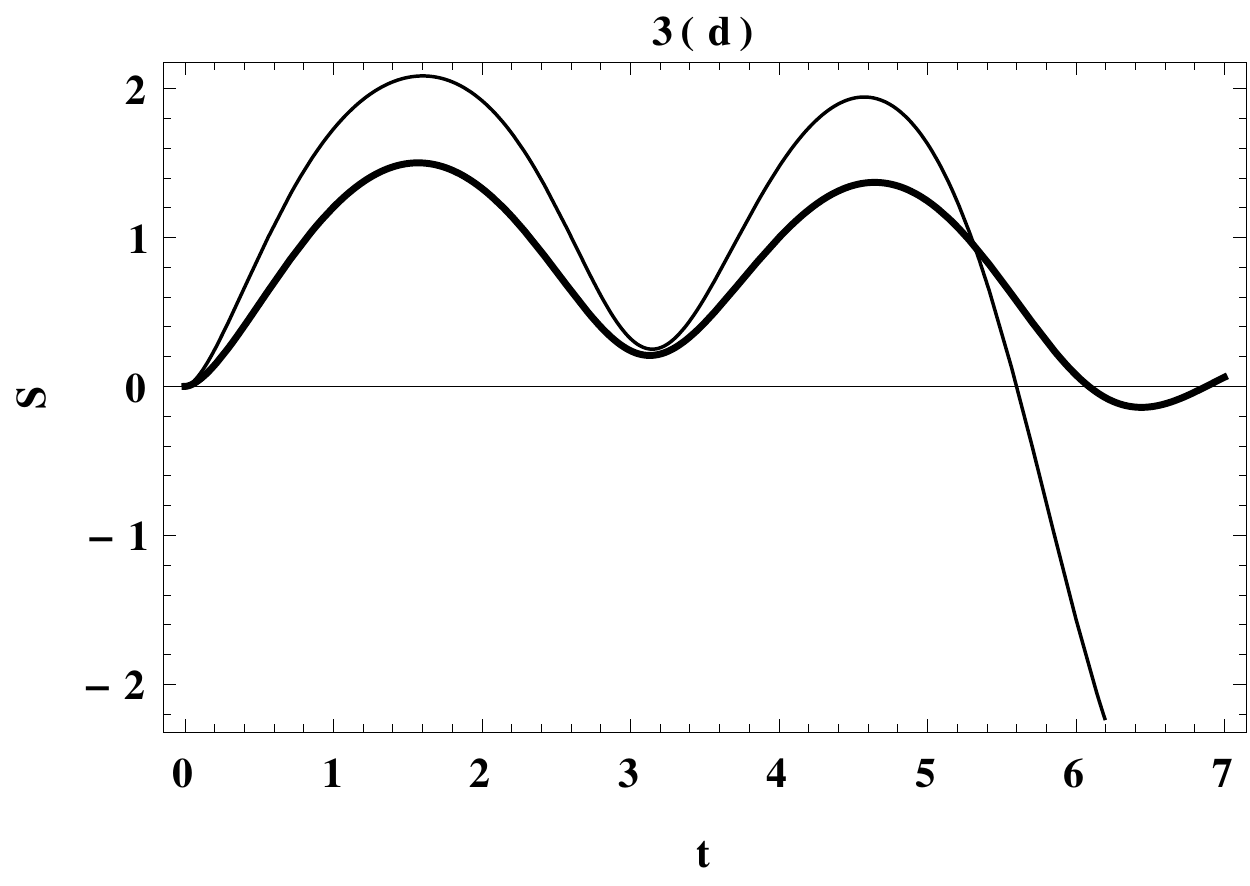} \\
\vspace{-0.5cm}
\tiny{\caption{The time evolution of entropy($S$) for two photons(shown in thick curve) and four photons(shown in thin curve):
fig(a) for $\gamma=0$(without damping), fig(b) for $\gamma = 0.01$,  fig(c) for $\gamma = 0.03$, fig(d) for $\gamma = 0.05$, with $J=0.5$.}}\label{75}
\end{figure}
\end{center} 

For a four photon state as an input state, the entropy of entanglement of the system,  

\begin{align} \nonumber  S  = &- 8 \cosh^8(\theta)\log[\cosh(\theta)]  - 8 \sinh^8(\theta)  \log[\sinh(\theta)] - 4 \sinh^2(\theta) \cosh^6(\theta)  \log[4 \sinh^2(\theta)  \cosh^6(\theta)] 
\\& - 6 \sinh^4(\theta) \cosh^4(\theta)  \log[6 \cosh^4(\theta) \sinh^4(\theta)]  - 4 \sinh^6(\theta) \cosh^2(\theta)  \log[4 \sinh^6(\theta)  \cosh^2(\theta)]   \end{align}

 is shown in thick curves of figs3(a,b,c,d). When $\gamma$ goes to zero, we get the same entropy in fig1(d) (without damping). 
As we increase the value of $\gamma$, we can see the damping effect, for four photon system there is  more damping than the  two photon system.
So, one can say that as we increase  the input photons, the system will decohere more. This can also be seen by calculating the decoherence parameter. 

Since the state is Gaussian, we can use the  the covariance matrix method to calculate the entanglement of the system by using Simon's criterion\cite{Simon}. 
The density matrix can be written as  $\rho^\prime(t)=S^\dag(r)R^\dag(\phi)\rho(t)R(\phi)S(r)$, 
where S(r) is the squeezing matrix and $R(\phi)$ is the rotation matrix mixing real and tilde fields. 
In our case,$\theta = 45^o$ (see in eqs(\ref{rot1}, \ref{rot2})) and  'r' is squeezing parameter(see in eqs(\ref{MN1}, \ref{MN2})), 
and $\rho(0)$ is the initial state of two mode system. 
Now we take the initial state $\rho(0)$ to be the two mode vacuum state, $|\rho(0)> = |0,0,\tilde{0},\tilde{0}>$. 
To calculate the entanglement of the time evolved state $\rho^\prime(t)$
we go over to phase space description by following transformations, \vspace{-0.5cm}

\be \nonumber A= \frac{1}{\sqrt{2}}(x+ip_x), \hspace{.5cm} A^\dag =  \frac{1}{\sqrt{2}}(x-ip_x), \hspace{.5cm} \tilde{A}= \frac{1}{\sqrt{2}}(\tilde{x}+i\tilde{p}_x), 
\hspace{.5cm} \tilde{A}^\dag =  \frac{1}{\sqrt{2}}(\tilde{x}-i\tilde{p}_x) \ee
\vspace{-0.1cm}
\be B= \frac{1}{\sqrt{2}}(y+ip_y), \hspace{.5cm} B^\dag =  \frac{1}{\sqrt{2}}(y-ip_y),  \hspace{.5cm} \tilde{B}= \frac{1}{\sqrt{2}}(\tilde{y}+i\tilde{p}_y),
\hspace{.5cm} \tilde{B}^\dag =  \frac{1}{\sqrt{2}}(\tilde{y}-i\tilde{p}_y). \ee

 Then, the covariance matrix is:
\be V(r_1,r_2) = \left(
{\begin{array}{cccccccc}
p & 0 & 0  & 0 & 0 & 0  & s & 0\\
0 & q & 0  & 0 & 0 & 0  & 0 & t \\
0 & 0 & p^*  & 0 & s & 0  & 0 & 0\\
0 & 0 & 0  & q^* & 0 & t  & 0 & 0\\
0 & 0 & s  & 0 & p & 0  & 0 & 0\\
0 & 0 & 0  & t & 0 & q  & 0 & 0\\
s & 0 & 0  & 0 & 0 & 0  & p^* & 0\\
0 & t & 0  & 0 & 0 & 0  & 0 & q^*\\
\end{array}}
\right) \label{cov} \ee

where, 
$p= e^{2iJt} e^{-2\gamma t}cosh 2r_1 $, $q=e^{2iJt} e^{-2\gamma t} cosh 2r_2 $, $s= e^{-2\gamma t}sinh 2r_1, $ $t=  e^{-2\gamma t}sinh 2r_2 $.

Since the tildian fields are fictitious, we trace over them to get  the covariance matrix for the physical modes, \vspace{-0.3cm}

\be V(r_1,r_2) = \left(
{\begin{array}{cccc}
p+q & 0 & -(s+t) & 0 \\
0 & p^*+q^* & 0  & s+t \\
-(s+t) & 0 & p+q  & 0 \\
0 & s+t & 0  & p^*+q^* \\
\end{array}}
\right) \ee

The canonical form of covariance matrix is given by, \vspace{-0.3cm}\be V = \left(
{\begin{array}{cc}
\alpha & \gamma  \\
\gamma^\dag & \beta \\
\end{array}}
\right) \ee

where, \be \alpha =  \left(
{\begin{array}{cc}
p+q & 0 \\
0 & p^*+q^* \\
\end{array}}
\right) = \beta, \hspace{.7cm}  and  \hspace{.7cm} \gamma =  \left(
{\begin{array}{cc}
-(s+t) & 0 \\
0 & (s+t) \\
\end{array}}
\right) \ee
Then the separablility condition\cite{Simon} for any two mode state is \vspace{-0.5cm}

\be Det\alpha Det\beta + (\frac{1}{4} - |Det\gamma|)^2 - tr(\alpha J \gamma J \beta J \gamma^T J) \ge \frac{1}{4} (Det\alpha + Det\beta) \label{simonrule} \ee

\begin{figure}
\includegraphics[width=2.7in,height=2in]{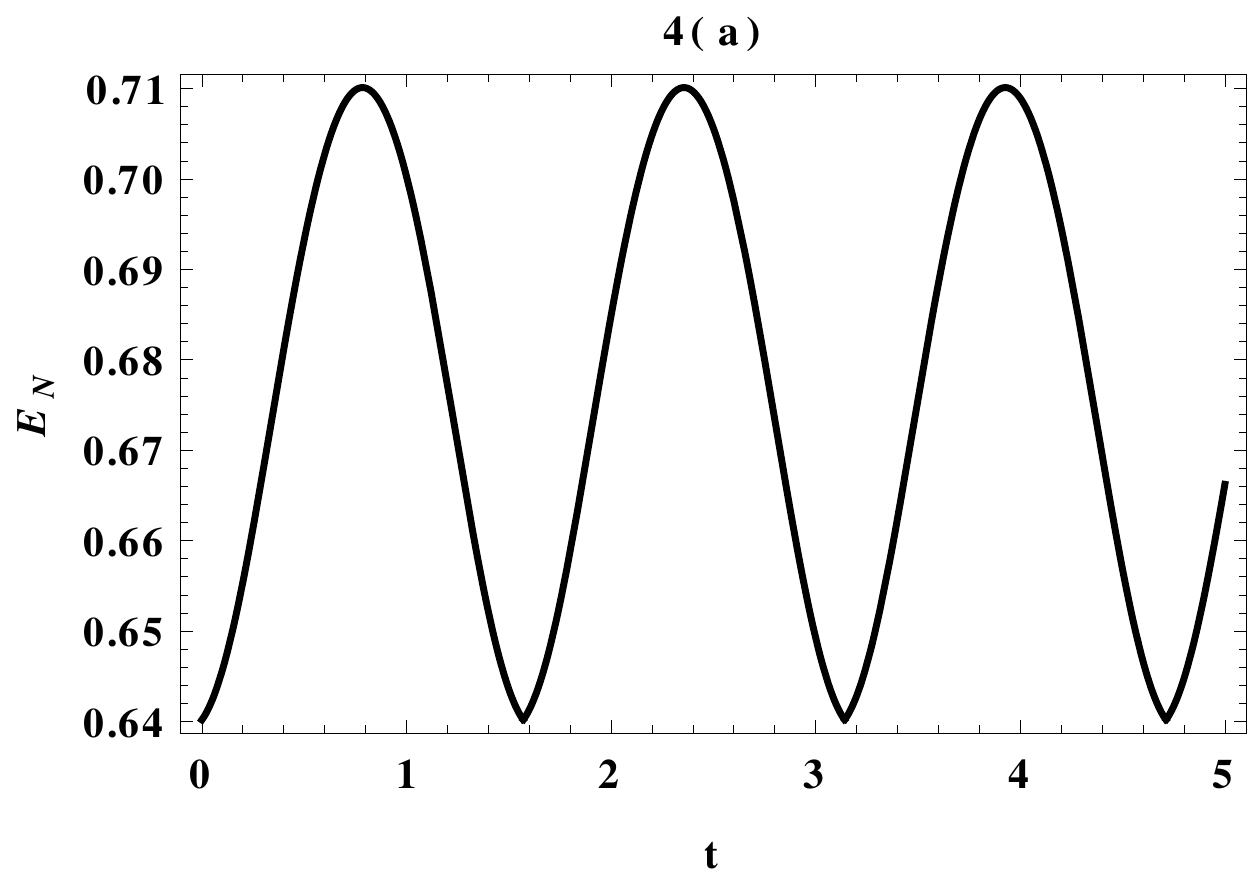} \hspace{0.5cm}  \includegraphics[width=3.3in,height=2in]{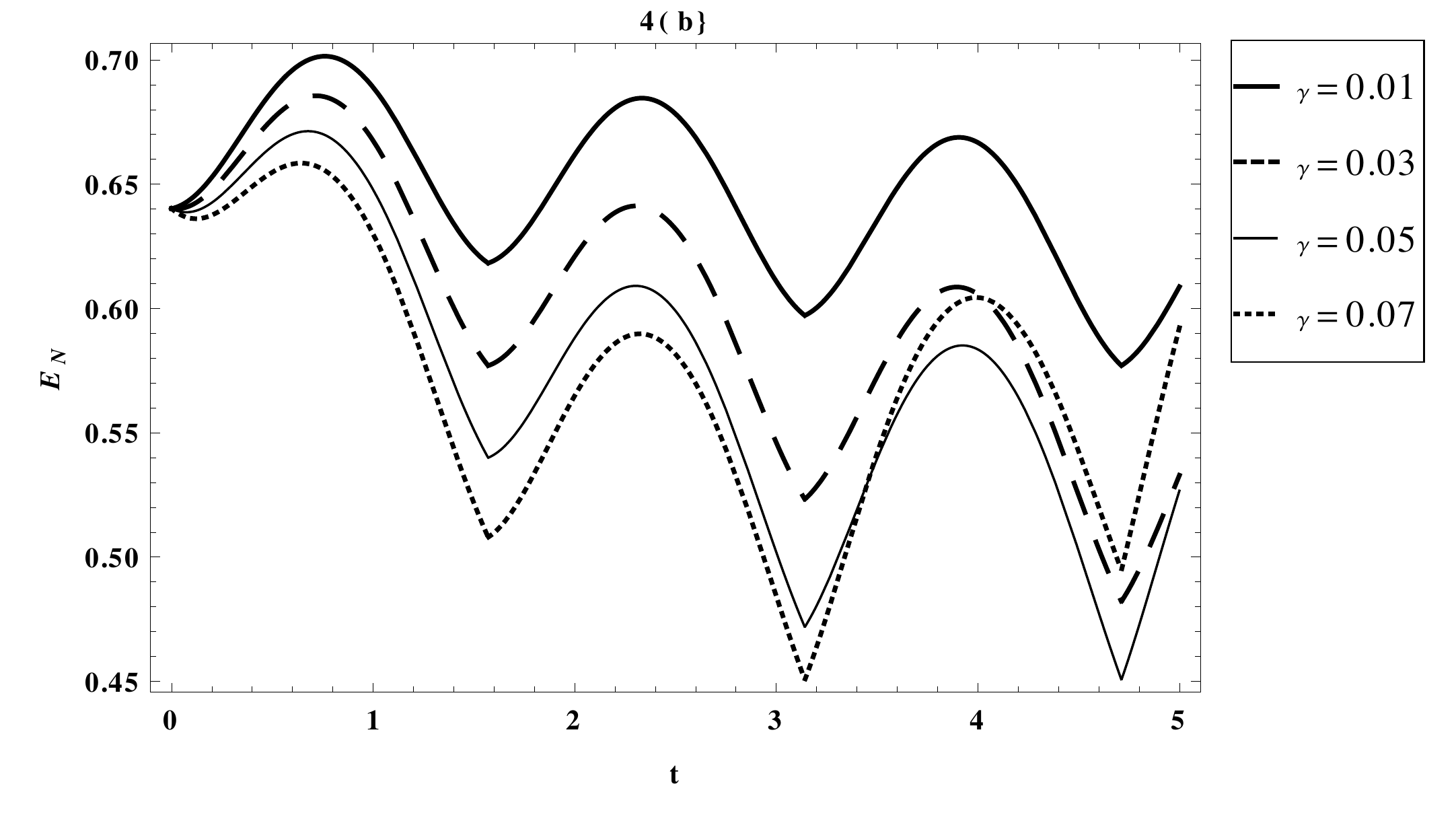} \\
\vspace{-0.5cm}
\tiny{\caption{(a) shows the time evolution of entanglement($E_N$) without damping($\gamma=0$), 
fig(b) shows the time evolution of entanglement with damping(for different values of $\gamma$) with $r=0.25, J=0.5$.}}\label{5}
\end{figure}

  The symplectic eigenvalues are defined as, \vspace{-0.3cm} \be \nu_\pm = \sqrt{\frac{1}{2}\{\tilde{\Delta} \pm \sqrt{\tilde{\Delta}^2 - \frac{4}{\mu^2}}\}} \ee
  where, 
$  \tilde{\Delta} = Det\alpha + Det\beta - 2Det\gamma  = 2(p+q)(p^*+q^*)+2(s+t)^2 $ \\ and
$ \mu = [Det V]^{-\frac{1}{2}} = [(p+q)^2(p^*+q^*)^2+(p+q)^2(s+t)^2-(p^*+q^*)^2(s+t)^2-(s+t)^4]^{-\frac{1}{2}} $
  
  The entanglement of the system is \be E_N = max\{0, -log\nu_-\}.  \label{QEn}\ee
  
  For $r_1 = r_2 = r$,    
   the entanglement for two mode vacuum states  {\bf without damping}$(i.e.,\gamma = 0)$ is shown in fig4(a) and
   {\bf with damping}$(i.e.,\gamma \ne 0)$ is shown in fig4(b). We see that as damping increases the entanglement decreases, but, for low damping,
   the system seems to sustain entanglement to a large extent, so that it is quite robust for applications.

In order to quantify  the  decoherence effects, 
we compute $\rho^2$  as 

\begin{align} \nonumber Tr[\rho^2(t)] & = Tr[\sum_{m,n} <m,n|\rho^2(t)|m,n>]
\\& =Exp\left[-\frac{4 \gamma t \sinh(\sqrt{2} \gamma +iJ)t}{ (\sqrt{2} \gamma +iJ)t\cosh(\sqrt{2} \gamma +iJ)t + (\gamma+iJt) \sinh(\sqrt{2} \gamma +iJ)t}\right] \end{align}

The behaviour of decoherence is plotted figs5.  We have considered two cases fig5(a),
shows the variation of decoherence with time for strong coupling  for various values of $\gamma$ and fig5(b),
shows the evolution of decoherence with weak  coupling. For strong  coupling, the system decoheres in an oscillatory fashion and saturates to a non-zero value,
while for weak coupling,  one that for even short times,  as the value of
damping coefficient increases the system decoheres, to a very low value, very fast.

\begin{center}
\begin{figure}
\includegraphics[width=3.15in,height=2in]{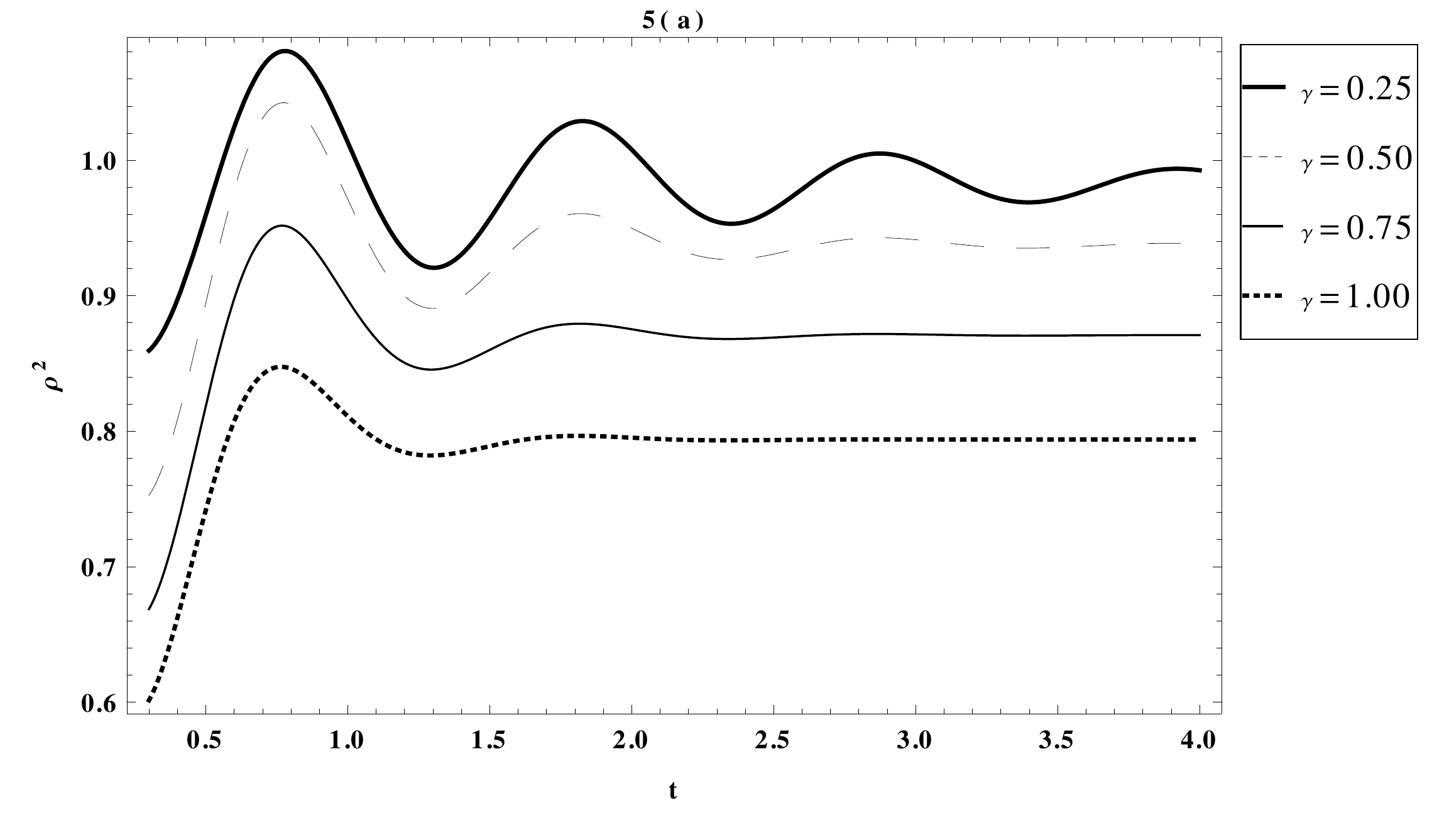} \hspace{0.1cm}  \includegraphics[width=2.9in,height=2in]{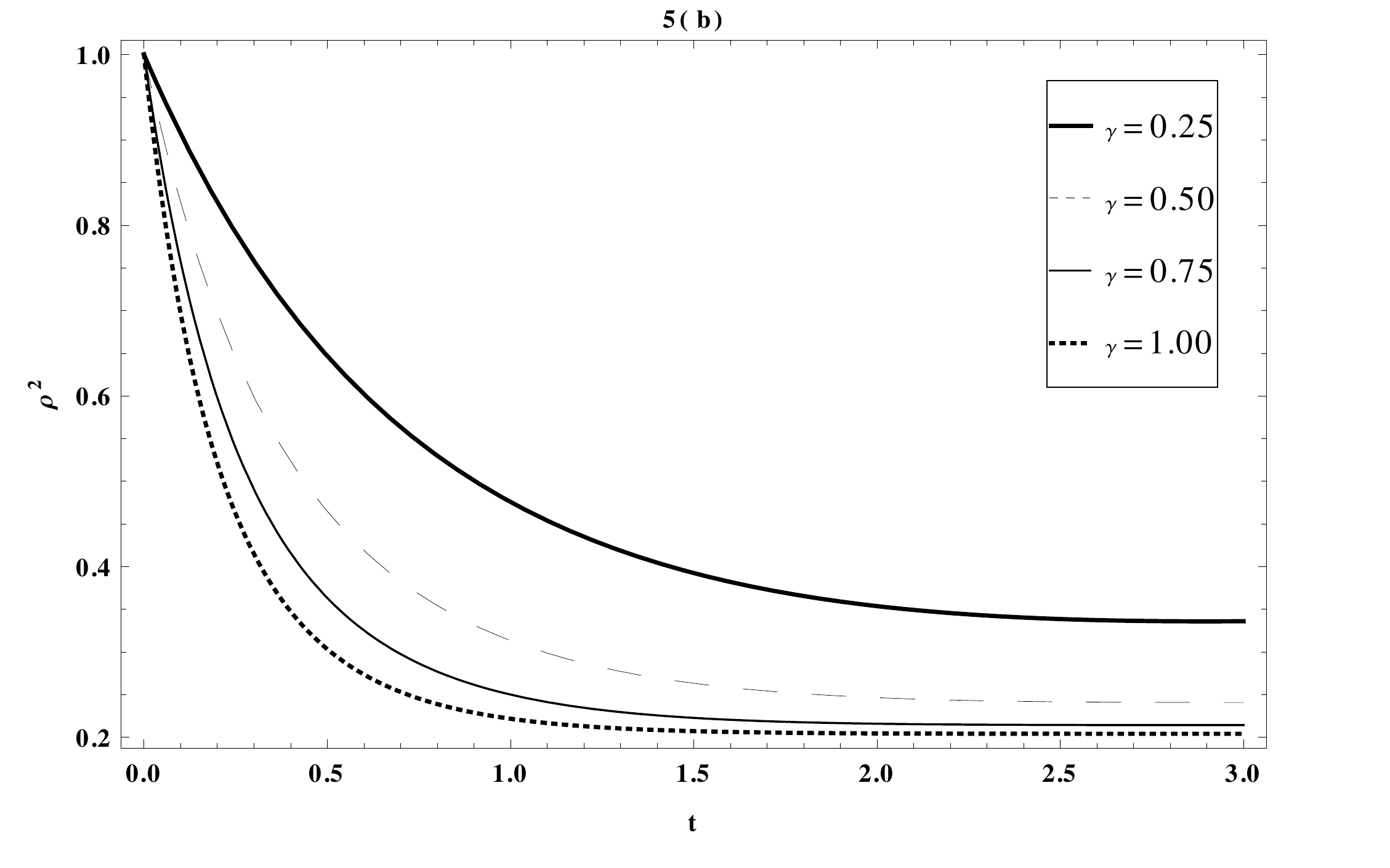} \\
\vspace{-0.7cm}
\tiny{\caption{(a) shows the time evolution of Decoherence for different values of $\gamma$ with $J=3$ and fig(b) for $J=0.25$ }}\label{14}
\end{figure}
\end{center}
 
 \subsubsection{Entanglement for two mode thermal state with damping $\gamma\ne0$}
 
Taking the initial state $\rho(0)$ to be the two mode thermal vacuum state, the  covariance matrix is given by,
\vspace{-0.5cm}
\be V(r_1,r_2) = \left(
{\begin{array}{cccc}
c+d & 0 & e+f & 0 \\
0 & c^*+d^* & 0  & -(e+f) \\
e+f & 0 & c+d & 0 \\
0 & -(e+f) & 0  & c^*+d^* \\
\end{array}}
\right) \ee

where,

$c=e^{2iJt} e^{-2\gamma t} ( n_1 cosh^2r_1 +n_2sinh^2r_1) $, $d= e^{2iJt} e^{-2\gamma t} (n_1sinh^2r_2+n_2cosh^2r_2)$,
$e=\frac{n_1+n_2}{2}e^{-2\gamma t}sinh 2r_1 $ and $f=\frac{n_1+n_2}{2}e^{-2\gamma t}sinh 2r_2 $.

Applying Simon's criterion  eq.(\ref{simonrule}) we see that the system is entangled iff
 \be (n_1+n_2)^4e^{-4\gamma t}[\cosh^22r-\sinh^22r )^2 +\frac{1}{16} \ge \frac{(n_1+n_2)^2}{2} e^{-2\gamma t}[\cosh^22r+\sinh^22r )\label{simonrule1} \ee
 For $r_1 = r_2 = r$, and $n_1 = n_2 =n$ , this condition is satisfied for the values of r given in the figure 5, 
 in which we plot the  logarithmic negativity as a function of $\bar{n}$, for different values of r. 
  We see that as the system not only gets less entangled for high values of $\gamma$ (quantified by r), but also for large n(external heat bath).
  So that in the presence of a heat bath the effect of damping increases and both have to be considered when generating entanglement in the lab by using coupled cavities.

\begin{figure}
\begin{center}
\includegraphics[width=4in,height=2.2in]{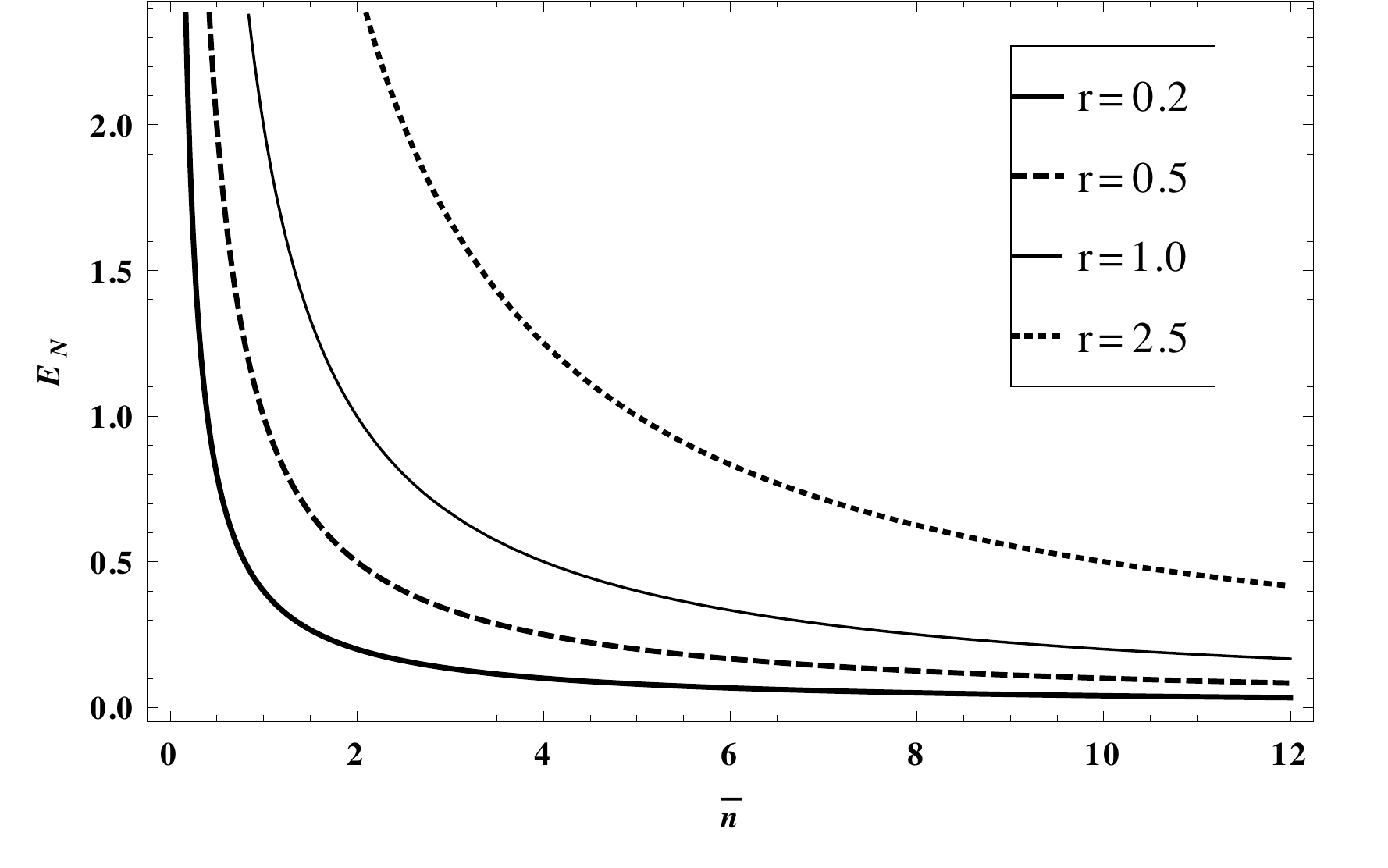}
\end{center}
\vspace{-0.7cm}
\tiny{\caption{Shows entanglement($E_N$)  vs. thermal distribution function($\bar{n}$) for different values of r }}\label{11}
\end{figure}

\section{Conclusion}
In this paper, we have shown that the formalism of thermofield dynamics is a powerful tool for exact studies of coupled waveguide systems. 
Indeed, we have exactly solved the master equation associated with SU(2) and SU(1,1) symmetries for coupled lossy waveguides with and without damping.
For coupled waveguides without damping, special attention has been given to the time evolution of the NOON states  as inputs and  we have
shown that as we increase the photon number, the entanglement of the NOON states survives with time, thus making them extremely  suitable for quantum information. 
The solution  for damped systems was obtained by transforming the master equation to a Schrodinger type equation and applying  the disentanglement formulae for SU(2)
and SU(1,1). Our work extends that of Rai et. al\cite{Agarwal}, as it gives the exact solution for the master equation, and,
in addition shows how the entanglement behaves for input thermal states.
Our results have also shown that the entanglement of the system can withstand a certain amount of damping, 
suggesting  that it can be used for applications such as quantum computation, even if the waveguides are lossy.
Furthermore we have shown the effect of an external heat bath on the system, by applying our methods to  thermal input states. 
 Our method shows the usefulness of thermofield dynamics in quantum entanglement problems, quite orthogonal to the approach given in ref.\cite{japanese},
 and allows us to handle damping in entanglement generation properties. We propose to apply this formalism to coupled light-atom systems,
 to shed further light on the effect of damping on the generation of entanglement. 

\section{Acknowledgements}
M.N.K. wishes to acknowledge CSIR-UGC for a JRF fellowship. KVSSC acknowledges the Department of Science and technology, Govt of India, (fast track
scheme (D. O. No: SR/FTP/PS-139/2012)) for financial support. We wish to thank Prof. C. Mukku and  Prof. S. Chaturvedi for insightful comments.
We also wish to thank V.Srinivasan for introducing us to  the details of thermofield dynamics.

\end{document}